\documentclass{article}
\usepackage{arxiv}
\usepackage[numbers,sort&compress]{natbib}
\usepackage[utf8]{inputenc} 
\usepackage[T1]{fontenc}    
\usepackage{ dsfont }
\usepackage{hyperref}       
\usepackage{url}
\usepackage{textcomp}
\usepackage{ gensymb }
\usepackage{amsmath}
\usepackage[ruled,vlined]{algorithm2e}
\usepackage{algorithmic}
\usepackage{booktabs}       
\usepackage{amsfonts}       
\usepackage{nicefrac}       
\usepackage{microtype}      
\usepackage{lipsum}
\usepackage{float}
\usepackage{graphicx}
\usepackage{multirow}
\usepackage{array}
\usepackage{comment}
\usepackage{longtable}
\usepackage{ragged2e}
\usepackage{subfig}

\title{Federated Learning in Healthcare: Model Misconducts, Security, Challenges, Applications, and Future Research Directions-A Systematic Review}

\author{
  Md Shahin Ali \\
  Department of Biomedical Engineering \\ Islamic University \\ Kushtia 7003, Bangladesh\\
  \texttt{shahinbme.iu@gmail.com} \\
   \And
  Md Manjurul Ahsan \\
  Department of Industrial and Systems Engineering\\
  University of Oklahoma\\
  Norman, Oklahoma-73071 \\
  \texttt{ahsan@ou.edu} \\
   \And
  Lamia Tasnim \\
  Department of Food and Nutrition \\ Government College of Applied Human Science \\ Azimpur, Dhaka 1205, Bangladesh\\
  \texttt{lamiafn.gcahs.du@gmail.com} \\
   \And
  Sadia Afrin \\
  Department of Food and Nutrition \\ Government College of Applied Human Science \\ Azimpur, Dhaka 1205, Bangladesh\\
  \texttt{sadiaafrin2271@gmail.com} \\
   \And
  Koushik Biswas \\
  Department of Computer Science and Engineering \\ IIIT Delhi, New Delhi \\ Delhi 110020, India\\
  \texttt{koushik313@gmail.com}\\
  \And
  Md Maruf Hossain \\
  Department of Biomedical Engineering \\ Islamic University \\ Kushtia 7003, Bangladesh\\
  \texttt{marufbme.iu@gmail.com} \\
   \And
  Md Mahfuz Ahmed \\
  Department of Biomedical Engineering \\ Islamic University \\ Kushtia 7003, Bangladesh\\
  \texttt{mahfuz.bme.iu@gmail.com} \\
   \And
  Ronok Hashan \\
  Department of Electrical and Electronic Engineering \\ Islamic University \\ Kushtia 7003, Bangladesh\\
  \texttt{ronokhasan8781@gmail.com} \\
   \And
  Md Khairul Islam \\
  Department of Biomedical Engineering \\ Islamic University \\ Kushtia 7003, Bangladesh\\
  \texttt{khairul.ice06@gmail.com} \\
   \And
  Shivakumar Raman \\
  Department of Industrial and Systems Engineering\\
  University of Oklahoma\\
  Norman, Oklahoma-73071\\
  \texttt{raman@ou.edu}} 

\begin{document}
\maketitle
\begin{abstract}
Data privacy has become a major concern in healthcare due to the increasing digitization of medical records and data-driven medical research. Protecting sensitive patient information from breaches and unauthorized access is critical, as such incidents can have severe legal and ethical complications. Federated Learning (FL) addresses this concern by enabling multiple healthcare institutions to collaboratively learn from decentralized data without sharing it. FL's scope in healthcare covers areas such as disease prediction, treatment customization, and clinical trial research. However, implementing FL poses challenges, including model convergence in non-IID (independent and identically distributed) data environments, communication overhead, and managing multi-institutional collaborations. A systematic review of FL in healthcare is necessary to evaluate how effectively FL can provide privacy while maintaining the integrity and usability of medical data analysis. In this study, we analyze existing literature on FL applications in healthcare. We explore the current state of model security practices, identify prevalent challenges, and discuss practical applications and their implications. Additionally, the review highlights promising future research directions to refine FL implementations, enhance data security protocols, and expand FL's use to broader healthcare applications, which will benefit future researchers and practitioners.

\end{abstract}


\maketitle

\section{Introduction}

\textbf{What is Federated Learning?} Federated Learning (FL) is a machine learning (ML) framework that enables clients spread across different geographical locations to collectively train a shared model without directly exchanging local data~\cite{yashwanth2023federated}. This framework tackles challenges like class imbalances and non-IID (non-independent and identically distributed) data distributions across clients. These issues can cause the ``client-drift" problem, resulting in slower convergence and suboptimal model performance~\cite{stripelis2023federated}. By periodically redistributing model updates and employing methods like adaptive self-distillation and data harmonization, FL improves model training while maintaining data privacy and security in distributed environments.

\textbf{What are the application domains of FL?} FL is becoming increasingly popular due to its effectiveness in training models while maintaining data security. There are several application domains where FL is making significant contributions. In healthcare, it enables hospitals to collaboratively analyze sensitive patient data without sharing it, improving diagnostic models~\cite{antunes2022federated}. Financial institutions leverage FL for fraud detection and credit scoring while ensuring customer data confidentiality\cite{long2020federated}. In smart devices, FL is used to personalize recommendations and predictive text on smartphones by training models locally. Autonomous vehicles benefit from FL by sharing model updates among fleets to enhance object detection and driving behavior predictions. In cybersecurity, FL aids in detecting anomalies across distributed networks, and in the Internet of Things (IoT), it facilitates device management and predictive maintenance~\cite{pokhrel2020federated}. Overall, FL's ability to train models securely makes it valuable across various industries where data privacy is crucial.

\begin{figure*}[ht]
    \centering
    \includegraphics[width=\textwidth]{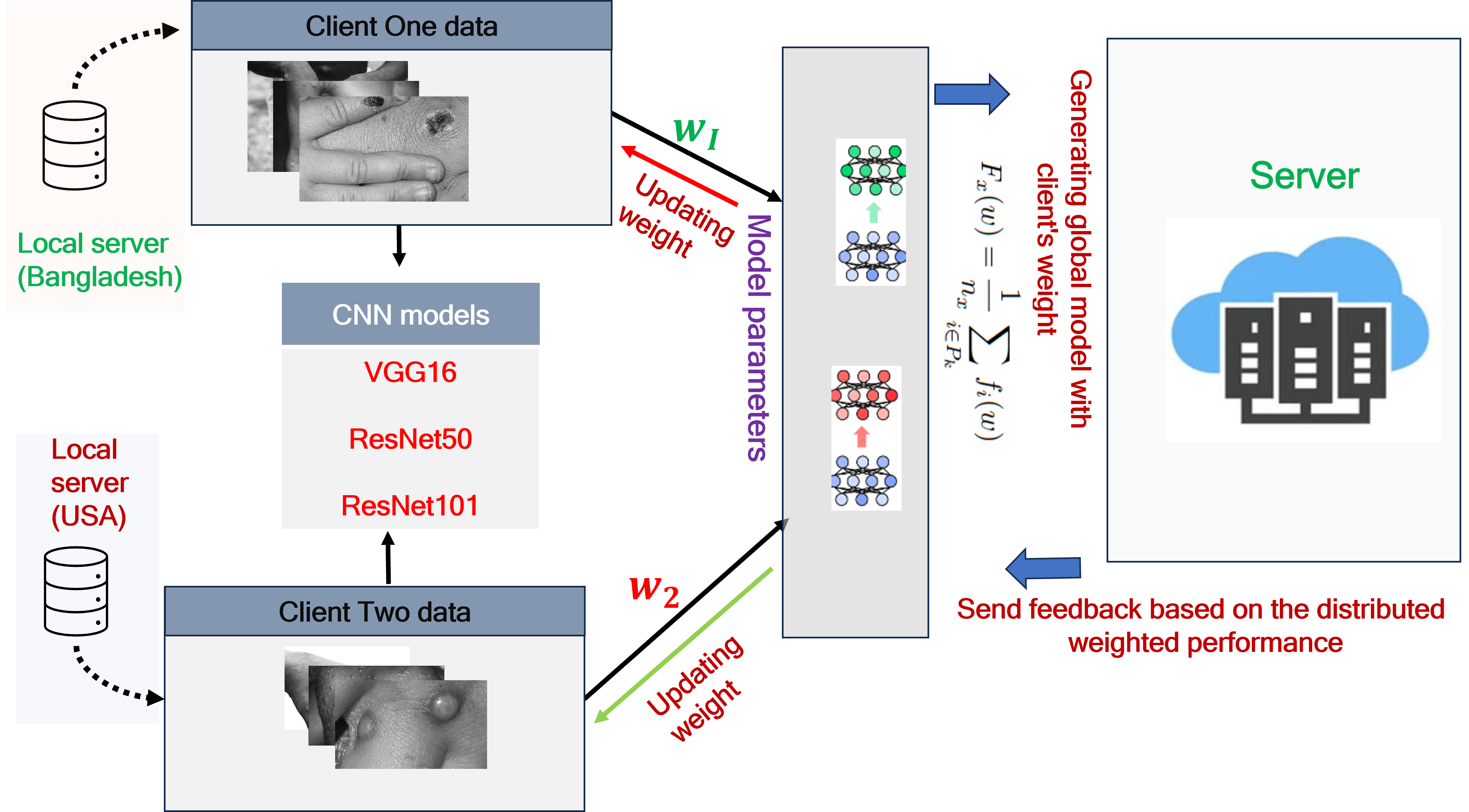}
    \caption{Privacy-focused FL for medical data: This distributed framework allows for the sharing of model weights among different clients. It assesses generalization on decentralized data by combining traditional deep learning with innovative \textbf{decentralized techniques}. This approach is used to improve predictive accuracy and provide detailed insights into the Monkeypox outbreak~\cite{ahsan2024enhancing}.}
    \label{fig:fed}
\end{figure*}

\textbf{What is FL in healthcare?}
FL is becoming popular in healthcare due to its application in data privacy, which is particularly important for handling sensitive patient medical data, such as images, reports, electronic health records (EHRs), and so on~\cite{antunes2022federated}. FL allows healthcare organizations to jointly train ML models while keeping raw data localized and secure, safeguarding patient privacy. This approach has been implemented across various healthcare sectors, particularly in disease diagnosis~\cite{ahsan2024enhancing}. Models trained on diverse, multi-institutional datasets have improved the accuracy of detecting conditions such as cancer, diabetic retinopathy, and other illnesses. 
Figure~\ref{fig:fed} illustrates the FL framework used to train various TL-based models, such as VGG16, ResNet50, and ResNet101, across two separate client servers. The clients exchange their trained weights, which serve as model parameters for the global model. 

Additionally, FL has enabled the secure sharing of EHR data, fostering the development of predictive models for patient outcomes and personalized treatment plans while maintaining privacy standards. In drug discovery, FL supports cross-institutional data collaboration, improving the identification of potential therapeutic compounds. Moreover, FL has facilitated the analysis of wearable device data, contributing to the early detection and monitoring of chronic diseases like heart disease and diabetes. By emphasizing privacy and security, FL encourages data collaboration while protecting sensitive medical information~\cite{xu2021federated}. Over the years, several key developments have significantly influenced the direction of FL in healthcare~\cite{nguyen2022federated,reddy2023comprehensive,moshawrab2023reviewing}:. 
\begin{itemize}
    \item \textbf{Privacy-Preserving Data Sharing:} FL enables healthcare organizations to collaborate on building robust ML models without sharing raw patient data. This is particularly important for sensitive medical information such as images, EHRs, and genomics data.  By keeping data localized and training models across multiple institutions, FL helps maintain patient privacy while leveraging a larger, more diverse dataset for improved model accuracy.
    \item \textbf{Enhancing Disease Diagnosis and Detection:} Federated models trained on diverse datasets from multiple healthcare institutions can detect diseases like cancer, diabetic retinopathy, and COVID-19 more accurately. They benefit from the shared expertise and data diversity across institutions. This collaborative approach leads to the early detection and better diagnosis of diseases, improving patient outcomes by providing clinicians with more reliable diagnostic tools.
\item \textbf{Optimizing Predictive Models for Patient Outcomes:} FL supports the development of predictive models that analyze patient histories and treatment plans to forecast patient outcomes and personalize treatments. Hospitals and clinics can use these predictive models to identify high-risk patients, adjust treatment plans proactively, and reduce hospital readmission's.

\item \textbf{Drug Discovery and Development:}
FL enables pharmaceutical companies and research institutions to collaboratively analyze data related to drug efficacy and safety without exposing proprietary information. This approach speeds up the identification of potential therapeutic compounds and allows for a more comprehensive assessment of drug effects across diverse patient populations.

\item \textbf{Wearable Device Data Analysis:} FL enables decentralized analysis of continuous data streams, helping clinicians monitor and predict chronic disease trends. Early detection and monitoring of conditions like heart disease and diabetes are significantly enhanced, leading to better patient management.

\end{itemize}

We organize this paper as follows: Section~\ref{2} discusses the methodology used in our study of FL in healthcare. Section~\ref{3} provides definitions, theories, and algorithms related to FL. Section~\ref{4} explores the types of FL. Section~\ref{5} examines misconduct threat models and adversarial goals in these systems. Section~\ref{6} looks at FL for healthcare applications. Section~\ref{7} addresses the challenges of implementing FL. Section~\ref{8} outlines potential future research directions. The paper concludes with Section~\ref{9}, which synthesizes our findings and suggests avenues for future research.

\begin{figure*}[ht]
    \centering
\includegraphics[width=\textwidth]{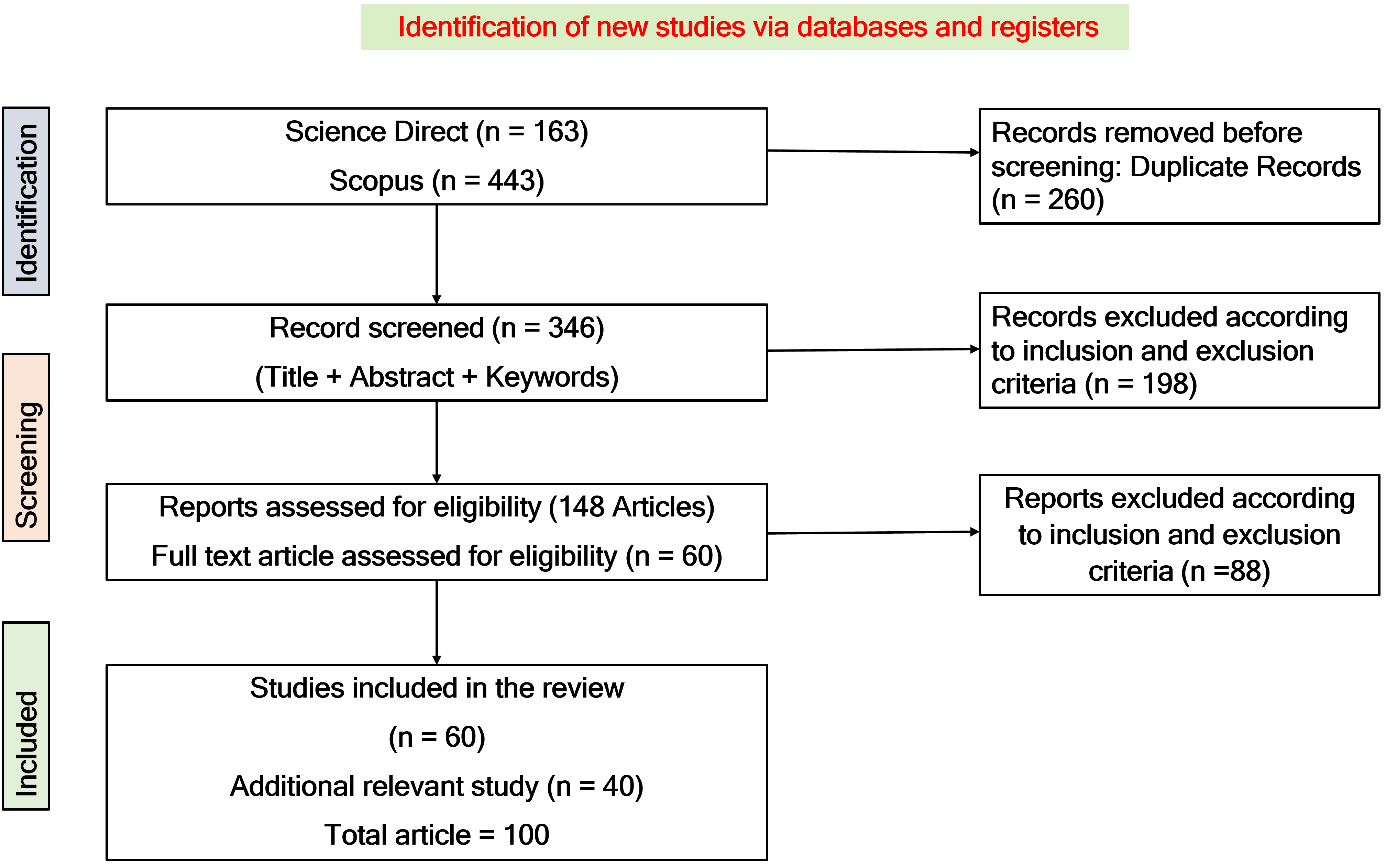}
    \caption{Study selection process depicted using the PRISMA flow diagram, including identification, screening, and inclusion steps.}
    \label{fig1111}
\end{figure*}

\section{Methodology}\label{2}
\begin{figure*}[ht]
    \centering
    \includegraphics[width=\textwidth]{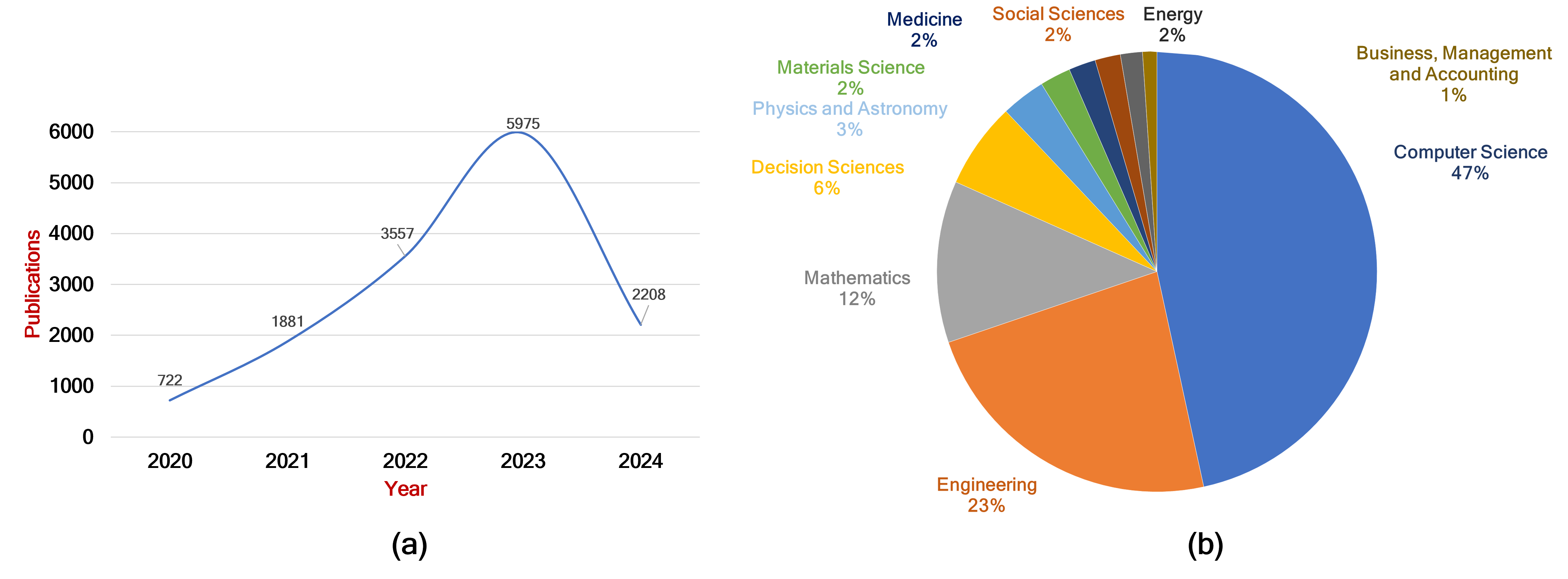}
    \caption{This survey examines the most recent advancements in FL within the field of computer science and engineering, particularly emphasizing the \textbf{\underline{healthcare}} sector. It presents statistics on (a) the number of papers published in the past five years on FL and (b) the distribution percentage of these papers across various domains. The plots illustrate a steady increase in recent literature.}
    \label{fig:pubyear}
\end{figure*}

This systematic review followed the Preferred Reporting Items for Systematic Reviews and Meta-Analyses (PRISMA) guidelines to ensure a transparent and comprehensive reporting of the review process. The guidelines were employed for identifying relevant studies, extracting data, and synthesizing findings. By adhering to the PRISMA guidelines, this review maintains methodological rigour, enhances reliability, and promotes the reproducibility of the systematic review~\cite{moher2009preferred}. The process of paper searching, screening, and selection followed the 2015 PRISMA guidelines, as shown in Figure~\ref{fig1111}. Initial searches were performed on two major databases, Scopus and Science Directory (SD) to find high-quality articles and SLRs from 2020 to 2024~\cite{ahsan2021machine,mustapha2021impact,manjurul2021machine,ahsan2022machine}. 
Figure~\ref{fig:pubyear} provides a statistical overview of papers published on FL in various disciplines over the last five years. Figure~\ref{fig:pubyear}(a) illustrates a substantial increase in research interest in FL from 2020 to 2023, with the number of publications rising from 722 to 5975. As of May 3, 2024, approximately 2208 papers have been published in FL domains this year, demonstrating ongoing progress across various fields. Figure~\ref{fig:pubyear}(b) shows that Computer Science is the predominant area, comprising 47\% of FL publications. It is followed by Engineering and Mathematics, which account for 23\% and 12\% of publications, respectively. In contrast, fields such as Medicine (2\%), Business, Management and Accounting (1\%), and Materials Science (2\%) have seen minimal exploration. These trends highlight a significant emphasis on the technical and computational elements of FL, while its potential in health, social sciences, and materials science remains relatively untapped.

For a more targeted and relevant search, Boolean operators were employed with the terms ``Federated Learning" and either ``healthcare," ``disease," or ``medical." This approach resulted in identifying 443 articles from Scopus and 163 articles from SD. Filtering for journal papers, peer reviews, and English language, the selection narrowed down to 163 articles after reviewing titles, keywords, and abstracts. One researcher (M.S. Ali) transferred data from these 163 journal articles into Excel CSV files for further detailed analysis. Using Excel’s tools to identify duplicates, redundant entries were removed. Titles and abstracts of these articles were independently reviewed by two reviewers (M.M. Ahsan and M.S. Ali). Any disagreements in article selection were resolved through discussion and consensus. Finally, following the inclusion and exclusion criteria depicted in Figure~\ref{fig1111}, a total of 100 papers were chosen for qualitative synthesis.

\section{Definitions, Theory, and Algorithms}
\textbf{Definitions}

FL is an ML framework designed to train a model across multiple decentralized edge devices or servers that hold local data samples, all while keeping the data localized. This approach helps preserve privacy and reduces the necessity of transferring large datasets. Mathematically, FL can be described as follows~\cite{yang2019federated}:

Let \( \mathcal{D}_i \) be the local dataset of the \( i \)-th participant in the federation, and \( f_i \) represent the local learning objective for that participant. The FL problem can be formulated as the optimization of a global objective function \( F \), which typically is an aggregation of the local objectives, which can be expressed as:

$$
\min_{\theta} F(\theta) = \min_{\theta} \sum_{i=1}^N w_i f_i(\theta; \mathcal{D}_i),
$$

where \( \theta \) represents the global model parameters, \( N \) is the number of participants, and \( w_i \) are the weights reflecting the relative importance or size of each local dataset \( \mathcal{D}_i \).

\vspace{3mm}

\textbf{Federated Learning in Healthcare}

The theoretical foundation of FL in healthcare is built upon the principles of distributed computing and privacy-preserving ML. In healthcare settings, where data is often fragmented across various institutions and contains sensitive patient information, FL offers a compelling solution. This approach allows for the collaborative training of ML models across multiple healthcare providers without requiring the exchange of raw data, thereby adhering to privacy regulations and ethical standards~\cite{singh2022framework}.

In practice, it can be implemented using various  ML techniques that support model training on decentralized data. Algorithms such as Stochastic Gradient Descent (SGD) are modified for the federated setting, where each participant computes gradients based on their local data~\cite{haddadpour2019convergence}. These gradients are then securely aggregated to update a shared global model. This model benefits from a diverse dataset encompassing different demographics and medical conditions, which enhances its ability to generalize and improves its diagnostic accuracy. For instance, an FL model could be trained to identify markers of disease across different populations using data from multiple hospitals, each contributing insights from their unique patient groups without compromising patient privacy~\cite{dayan2021federated}.

\vspace{3mm}
\textbf{General View of the Algorithm}

We provide a detailed overview of the algorithmic aspects of FL for healthcare applications. This overview contains several key steps, such as data partitioning, local model training, and global model aggregation. A general overview of the typical steps involved in implementing FL in healthcare is presented below~\cite{mothukuri2021survey}:


\vspace{3mm}

\textbf{Input:} Training datasets \( D_1, D_2, \ldots, D_m \) from \( m \) different healthcare institutions, Global model \( M \), Number of communication rounds \( R \), Local training epochs \( E \), Learning rate \( \eta \).

\vspace{3mm}

\textbf{Output:} Trained global model \( M \)

\underline{Data Partitioning:}
\begin{enumerate}
    \item Partition the overall training data into \( m \) subsets, \( D_1, D_2, \ldots, D_m \), where \( D_i \) represents the local dataset of the \( i \)-th healthcare institution.
\end{enumerate}

\underline{Local Model Training:}
\begin{enumerate}
    \item For each communication round \( r \) from \( 1 \) to \( R \):
    \begin{enumerate}
        \item Send the global model \( M^{(r-1)} \) to all \( m \) institutions.
        \item Each institution \( i \) initializes its local model \( M_i^{(r)} \) with \( M^{(r-1)} \).
        \item Each institution \( i \) trains \( M_i^{(r)} \) on its local dataset \( D_i \) for \( E \) epochs using a local optimizer with learning rate \( \eta \), resulting in updated model \( M_i^{(r,E)} \).
    \end{enumerate}
\end{enumerate}

\underline{Global Model Aggregation:}
\begin{enumerate}
    \item Collect the updated models \( M_1^{(r,E)}, M_2^{(r,E)}, \ldots, M_m^{(r,E)} \) from all \( m \) institutions.
    \item Aggregate the local models to update the global model:
    \[
    M^{(r)} = \frac{1}{m} \sum_{i=1}^{m} M_i^{(r,E)}
    \]
\end{enumerate}

\underline{Prediction:}
\begin{enumerate}
    \item Given a new example \( x \), use the trained global model \( M^{(R)} \) to predict the label \( y \).
    \item Return the predicted label \( y \).
\end{enumerate}



\section{Types of Federated Learning}\label{3}


\subsubsection{Horizontal Federated Learning}
Horizontal Federated Learning (HFL) represents a distributed ML paradigm aimed at enhancing model performance by harnessing data from decentralized devices or nodes. This technique, often applied in scenarios involving Internet of Things (IoT) devices or mobile devices, enables collaborative learning while preserving data privacy and security. HFL operates through horizontal federation, where data is partitioned horizontally across multiple nodes, allowing each node to process its portion of the data locally. This decentralized approach reduces the risk of data breaches and ensures that sensitive data remains on users' devices, addressing significant concerns regarding privacy and data security \cite{shi2022deep, lim2020federated}.

\begin{figure}[ht]
    \centering
\includegraphics[width=90mm, height= 75mm]{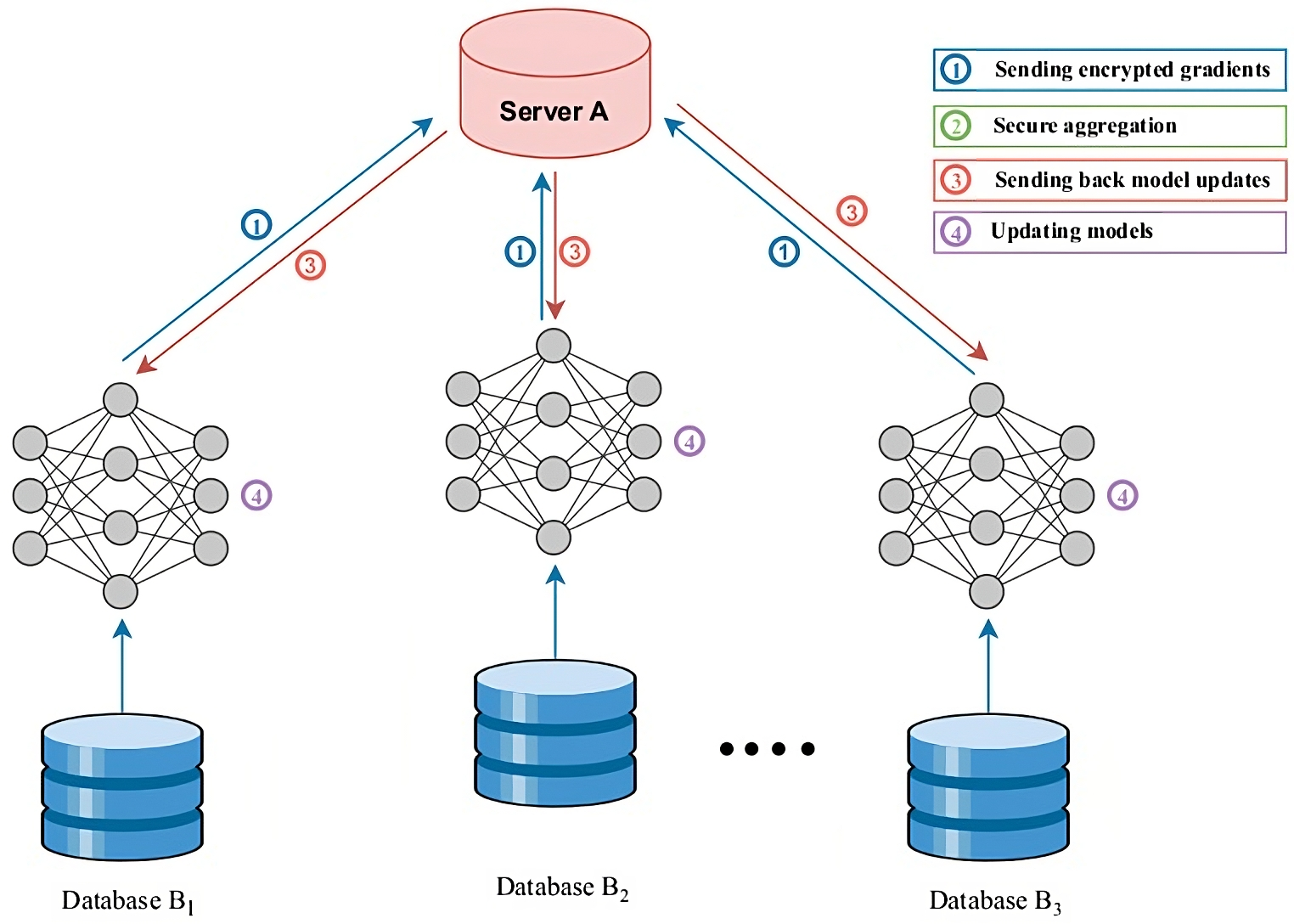}
    \caption{Horizontal FL in the healthcare sector \cite{yang2019federated}.}
    \label{fig-2}
\end{figure}

A notable advantage of HFL lies in its ability to leverage a larger and more diverse dataset compared to centralized ML methods. By aggregating data from multiple nodes, HFL enables models to be trained on a comprehensive range of data sources, leading to improved accuracy and robustness \cite{wang2019measure}. This distributed learning approach also offers potential benefits in various domains, including healthcare, where it allows for collaborative training on electronic health records (EHR) from multiple institutions without compromising patient privacy \cite{shi2022deep}.

However, despite its advantages, HFL presents several challenges that need to be addressed. Ensuring data consistency and quality across distributed nodes is a critical concern, as biases and errors in individual node data can significantly impact model performance \cite{suzen2020novel}. Strategies such as data preprocessing and quality checks are employed to mitigate these challenges and ensure the reliability of the training process \cite{zhang2021survey}.

\begin{table*}[]
\caption{Uses of Horizontal FL in the healthcare sector.}
\label{table-2}
\centering
\scalebox{.9}{
\begin{tabular}{|c|c|}
\hline
Application & Related Research Papers \\ \hline
Heart Disease Prediction & \begin{tabular}[c]{@{}c@{}}Hybrid Classifier-Based FL in Health Service   Providers \\ for Cardiovascular Disease Prediction \cite{yaqoob2023hybrid}\end{tabular} \\ \hline
Mortality Prediction & \begin{tabular}[c]{@{}c@{}}FL of electronic health records to improve mortality prediction \\ in hospitalized patients with COVID-19: machine learning approach \cite{vaid2021federated}   \\ \end{tabular} \\ \hline
Image Analysis &  \\ \hline
MRI Image Analysis & \begin{tabular}[c]{@{}c@{}}FL of generative image priors for MRI reconstruction    \cite{elmas2022federated} \\ \end{tabular} \\ \hline
CT Image Analysis & \begin{tabular}[c]{@{}c@{}}Blockchain-federated learning and deep learning models for \\ covid-19   detection using ct imaging \cite{kumar2021blockchain} \end{tabular} \\ \hline
fMRI Image Analysis & \begin{tabular}[c]{@{}c@{}}Multi-site fMRI analysis using privacy-preserving FL   \\ and domain adaptation: ABIDE results \cite{li2020multi} \end{tabular} \\ \hline
Disease Detection &  \\ \hline
COVID-19 Detection & \begin{tabular}[c]{@{}c@{}}Dynamic-fusion-based federated learning for  COVID-19 detection \cite{zhang2021dynamic} \end{tabular} \\ \hline
Skin Cancer Detection & \begin{tabular}[c]{@{}c@{}}Multimodal melanoma detection with FL \cite{agbley2021multimodal} \end{tabular} \\ \hline
Clinical Decision Support &  \\ \hline
\begin{tabular}[c]{@{}c@{}}Diabetic Retinopathy \\ Decision Support\end{tabular} & \begin{tabular}[c]{@{}c@{}}Making FL robust to adversarial attacks by learning \\ data and model association \cite{qayyum2022making}\end{tabular} \\ \hline
\begin{tabular}[c]{@{}c@{}} Heart Failure Decision \\ Support\end{tabular} & \begin{tabular}[c]{@{}c@{}}FedSDM: FL based smart decision making module for ECG data in \\ IoT integrated Edge-Fog-Cloud computing environments \cite{rajagopal2023fedsdm}\end{tabular} \\ \hline
Natural Language Processing &  \\ \hline
\begin{tabular}[c]{@{}c@{}}Electronic Health Record \\ Text Analysis\end{tabular} & \begin{tabular}[c]{@{}c@{}}Protecting personal healthcare record using blockchain \&  FL technologies \cite{aich2022protecting}\end{tabular} \\ \hline
\end{tabular}}
\end{table*}

Efficient and scalable implementation of the training process remains another significant challenge in HFL. Training on combined data from multiple nodes requires optimization and scaling strategies, including parallel processing and distributed training, to manage the resource-intensive nature of the task \cite{lu2022personalized, chamikara2021privacy}. Despite these challenges, HFL offers a promising approach to improve model accuracy while preserving privacy in distributed environments. HFL in healthcare systems is illustrated in Figure \ref{fig-2}, showcasing its application and workflow within the domain. Additionally, Table \ref{table-2} demonstrates the diverse applications of HFL in healthcare, particularly in predictive modeling. The table highlights insights from various studies, providing a comprehensive overview of authors' perspectives and methodologies utilized in implementing HFL in healthcare systems.

\subsubsection{Vertical Federated Learning}

Vertical Federated Learning (VFL) aims to enhance the efficacy of ML models by leveraging data from diverse distributed nodes or devices \cite{li2020review}. This method involves vertically partitioning the data of each node, ensuring that each node possesses a distinct set of attributes or characteristics, and subsequently training the model using collective data from all nodes. The proposed approach aligns with findings reported by Pfitzner et al. \cite{pfitzner2021federated}.

\begin{figure*}[ht]
    \centering
\includegraphics[width=\linewidth]{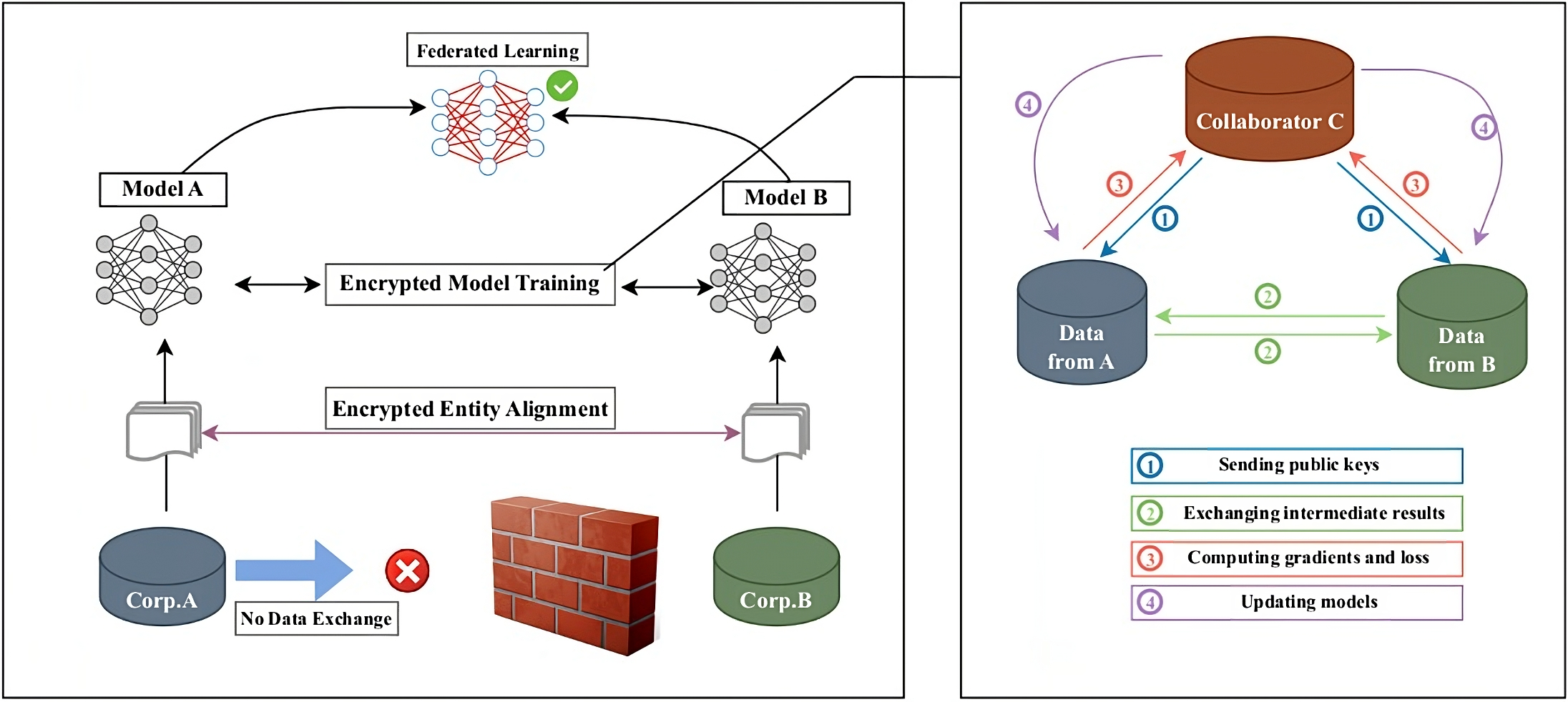}
    \caption{Vertical FL in the healthcare sector \cite{wang2020fed}.}
    \label{fig-3}
\end{figure*}

VFL finds widespread application in domains such as healthcare and finance, where heterogeneous data is commonly gathered from diverse devices and nodes \cite{sui2020feded}. In healthcare, VFL enables the aggregation of medical records from numerous patients, contributing unique information for model training \cite{oh2023federated}. This study proposes a hybrid ML approach for medical diagnosis, training the model on data from all patients to expand the dataset significantly compared to centralized ML approaches \cite{yaqoob2023hybrid}. The application of VFL in the healthcare system is visually depicted in Figure \ref{fig-3}.

The pre-processing phase of VFL involves vertical data partitioning, dividing data from each node into distinct subsets corresponding to specific data categories \cite{lau2021risk}. Subsequently, the data is introduced into the model for training \cite{varasteh2023privacy}, enhancing overall performance by learning from a larger and more varied dataset \cite{lavaur2022evolution}.

Ensuring consistency and high-quality performance across all nodes poses a significant challenge in VFL \cite{gandhi2021federated}. To address this challenge, VFL typically implements quality control measures and data refinement techniques \cite{thapa2022splitfed}.

The practicality and scalability of training in VFL is a recognized challenge \cite{thapa2022splitfed,yao2021analysis}. Optimization and scaling approaches, such as parallel processing and distributed training, are commonly employed to overcome these challenges. The study presents equations for VFL derived from the optimization problem of minimizing the loss function while preserving data privacy~\cite{chamikara2021privacy}.

\begin{table*}[]
\caption{Vertical FL and its uses in the Healthcare sector.}
\label{table-3}
\scalebox{.87}{
\begin{tabular}{|ll|l|}
\hline
\multicolumn{2}{|l|}{\textbf{Application}} & \textbf{Related Research Papers} \\ \hline
\multicolumn{1}{|l|}{\multirow{3}{*}{\textbf{Disease Diagnosis}}} & Heart Disease Diagnosis & \begin{tabular}[c]{@{}l@{}}Communication-efficient vertical FL    \cite{khan2022communication}\end{tabular} \\ \cline{2-3} 
\multicolumn{1}{|l|}{} &  Cancer Diagnosis & \begin{tabular}[c]{@{}l@{}}A review of medical FL: Applications in\\ oncology and cancer research \cite{chowdhury2022review}\end{tabular} \\ \cline{2-3} 
\multicolumn{1}{|l|}{} & COVID-19 Diagnosis & \begin{tabular}[c]{@{}l@{}}Covid-19 imaging data privacy by FL design: \\ A   theoretical framework \cite{ ulhaq2020covid}\end{tabular} \\ \hline
\multicolumn{1}{|l|}{\multirow{2}{*}{\textbf{Clinical Decision Support}}} & Predicting Adverse Drug Reactions & \begin{tabular}[c]{@{}l@{}}FL for healthcare domain-Pipeline, applications\\ and   challenges \cite{joshi2022federated}. \end{tabular} \\ \cline{2-3} 
\multicolumn{1}{|l|}{} & Diagnosis Support for Rare Diseases & \begin{tabular}[c]{@{}l@{}}FL for healthcare domain-Pipeline, applications \\ and   challenges. \cite{joshi2022federated}. \end{tabular} \\ \hline
\multicolumn{1}{|l|}{\textbf{Electronic Health Record   Analysis}} & Clinical Outcome Prediction & \begin{tabular}[c]{@{}l@{}}FL approaches for fuzzy cognitive maps to support   \\ clinical decision-making in dengue \cite{ hoyos2023federated}. \end{tabular} \\ \hline
\multicolumn{1}{|l|}{\multirow{2}{*}{\textbf{Imaging Analysis}}} & MRI Image Segmentation & \begin{tabular}[c]{@{}l@{}}Review on security of FL and its application in   \\ healthcare \cite{ li2023review}. \end{tabular} \\ \cline{2-3} 
\multicolumn{1}{|l|}{} & PET Image Analysis & \begin{tabular}[c]{@{}l@{}}A multi‐modal heterogeneous data mining algorithm using \\ federated learning \cite{ wei2021multi}. \end{tabular} \\ \hline
\end{tabular}}
\end{table*}

Vertical FL represents a distributed ML technique that leverages collective data from decentralized devices or nodes, enabling the development of more precise models while safeguarding user privacy \cite{wu2023topology,ouyang2023artificial}. This methodology allows for training on a significantly more extensive and diversified dataset compared to conventional centralized ML approaches \cite{lo2021systematic}, offering promising implications for various domains, including healthcare and finance. Table \ref{table-3} illustrates the diverse applications of VFL in healthcare, showcasing its predictive modeling capabilities alongside insights from various studies on the topic.

\subsubsection{Federated Transfer Learning}

The objective of Federated Transfer Learning (FTL), an ML technique, is to ensure confidentiality while transferring information from a centralized model to a decentralized one \cite{chen2020fedhealth}. Fan et al., in their work on IoT Defender \cite{fan2020iotdefender}, suggest an approach where a centralized model is trained first, followed by the transmission of acquired knowledge to decentralized models trained on smaller datasets. That approach aims to enhance the performance and precision of distributed models by assimilating knowledge from the centralized model.

\begin{figure*}[ht]
    \centering
\includegraphics[width=120mm,height=80mm]{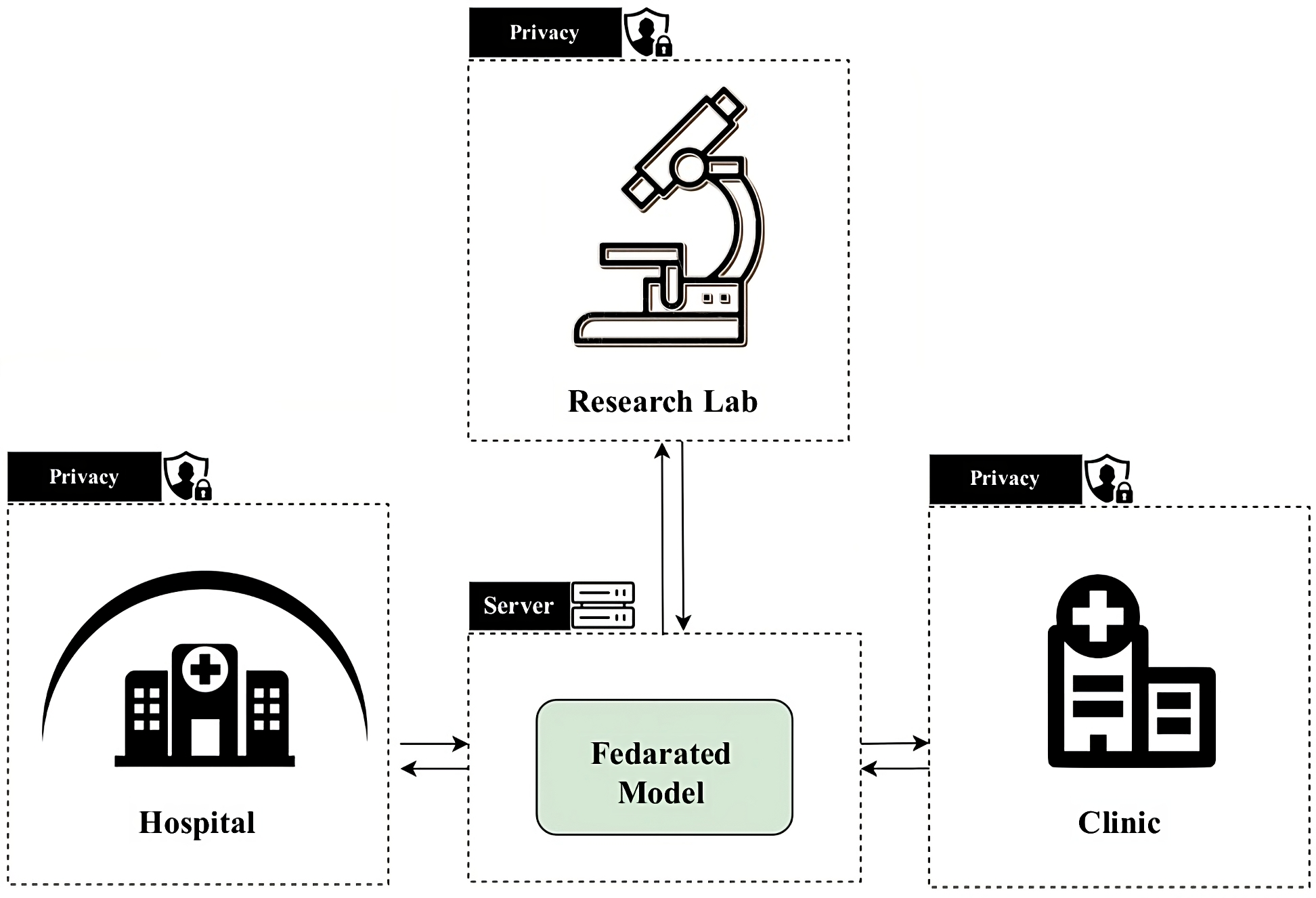}
    \caption{Federated Transfer Learning in the healthcare sector \cite{joshi2022federated}.}
    \label{fig-4}
\end{figure*}

FTL has emerged as a promising technique for training more precise models while preserving individual privacy \cite{li2021federated}. By leveraging knowledge learned from one domain to another, FTL improves model performance in new domains while safeguarding individual data privacy. This approach is particularly relevant in scenarios with sensitive or distributed data, minimizing the risk of data breaches and unauthorized access \cite{sun2022federated}. The application of FTL has become prevalent in IoT and mobile device environments, addressing limitations posed by distributed nodes or devices with limited resources. This approach facilitates secure and privacy-preserving learning processes by eliminating the need for centralized data storage \cite{rieke2020future}.

Zheng, Xiao, et al. explored the efficacy of FTL in knowledge transfer across domains, utilizing strategies such as fine-tuning, knowledge distillation, and model compression \cite{zheng2021mobile}. In other studies, the authors examine the relationship between knowledge distillation and model compression, both involving the central model, to facilitate training of distributed models and enable local training \cite{chen2021fedhealth,can2021privacy}. Additionally, Zhang, Wei, et al. present a comprehensive analysis of equations for FTL, exploring the mathematical formulations governing knowledge transfer in federated learning. Their analysis underscores the importance of problem-specific settings and the nature of data in FTL equations \cite{zhang2022data}.

Figure \ref{fig-4} visually illustrates the application of FTL in healthcare, providing insights into its functionality and operation within the healthcare domain.

\subsubsection{Federated Domain Adaption}

\begin{figure*}[ht]
    \centering
\includegraphics[width=110mm,height=85mm]{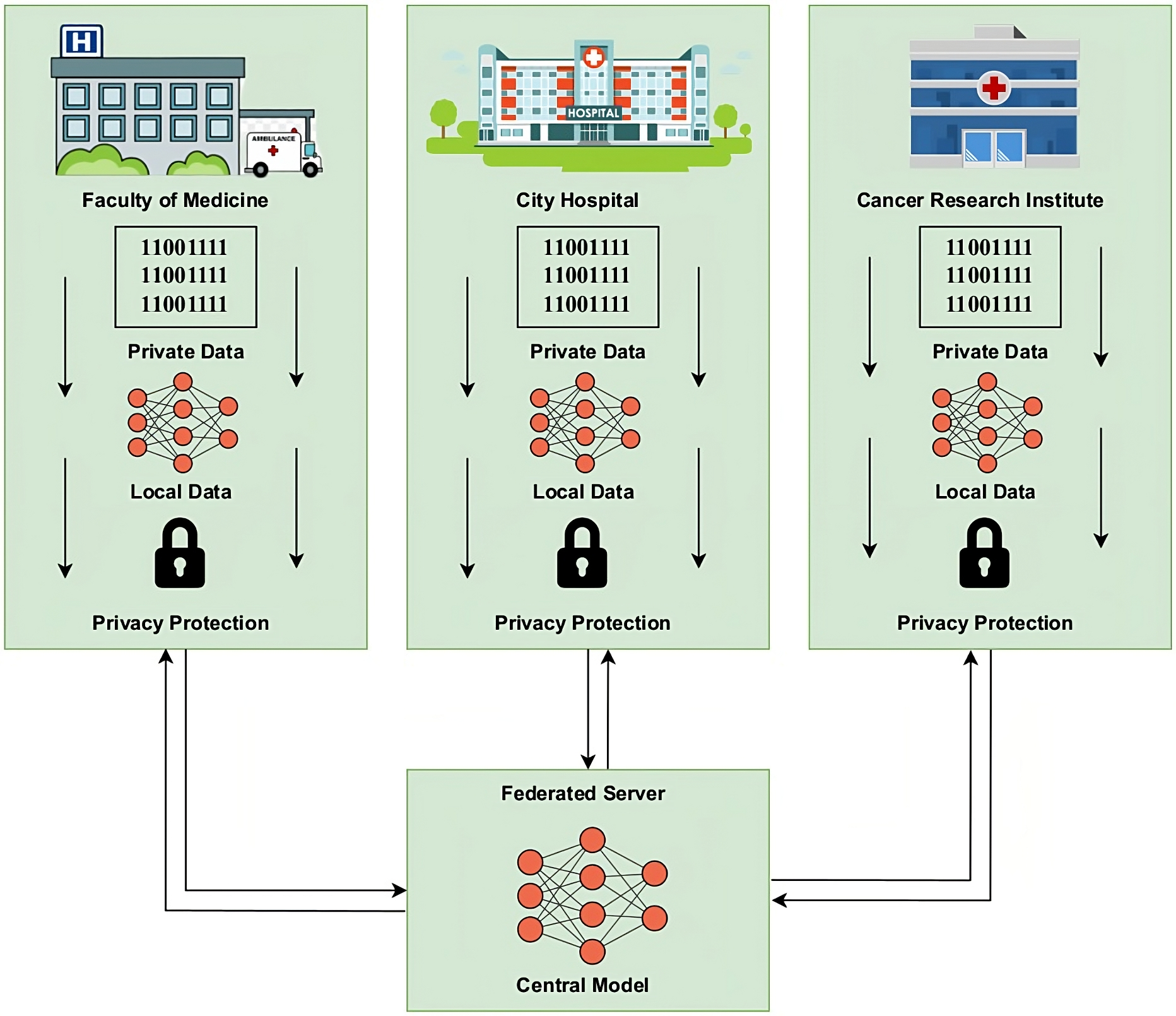}
    \caption{Federated domain adaption in the healthcare sector~\cite{suzen2020novel}.}
    \label{fig-5}
\end{figure*}

Federated Domain Adaptation (FDA) is a well-established ML technique widely applied across various domains. FDA facilitates knowledge transfer between domains while safeguarding user privacy and addressing the challenge of domain shift. The method has been extensively documented in previous research, demonstrating promising results, particularly in smart systems \cite{zeadally2020smart}. Leveraging domain adaptation algorithms, FDA aligns distributions between a source domain, typically data-rich, and a target domain, often characterized by limited data and annotations \cite{appari2010information}. By reducing the divergence between domains, the FDA enhances model generalizability, especially in scenarios with sparse target domain data \cite{abdel2022collaborative}. Figure \ref{fig-5} visually illustrates the FDA's application in healthcare, providing insights into its functionality within this domain.

Utilized where data centralization is impractical or undesirable, FDA enables the development of precise models while preserving data confidentiality \cite{griebel2015scoping}. This approach strategically utilizes existing data to train models suitable for target domains with limited data and distinct distributions from source domains \cite{kontar2021internet}.

The adaptation process involves adjusting a pre-trained model from the source domain to fit the target domain's data distribution, ensuring sensitive data remains on users' devices \cite{ratnapalan2020health}. The updated model parameters are then merged to create a global model tailored for the target domain \cite{glick2016new}. The FDA facilitates knowledge transfer across domains while preserving privacy, benefiting scenarios with limited target domain data and distinct distributions.

\subsubsection{Multitask Federated Learning}

\begin{figure*}[ht]
    \centering
\includegraphics[width=100mm,height=80mm]{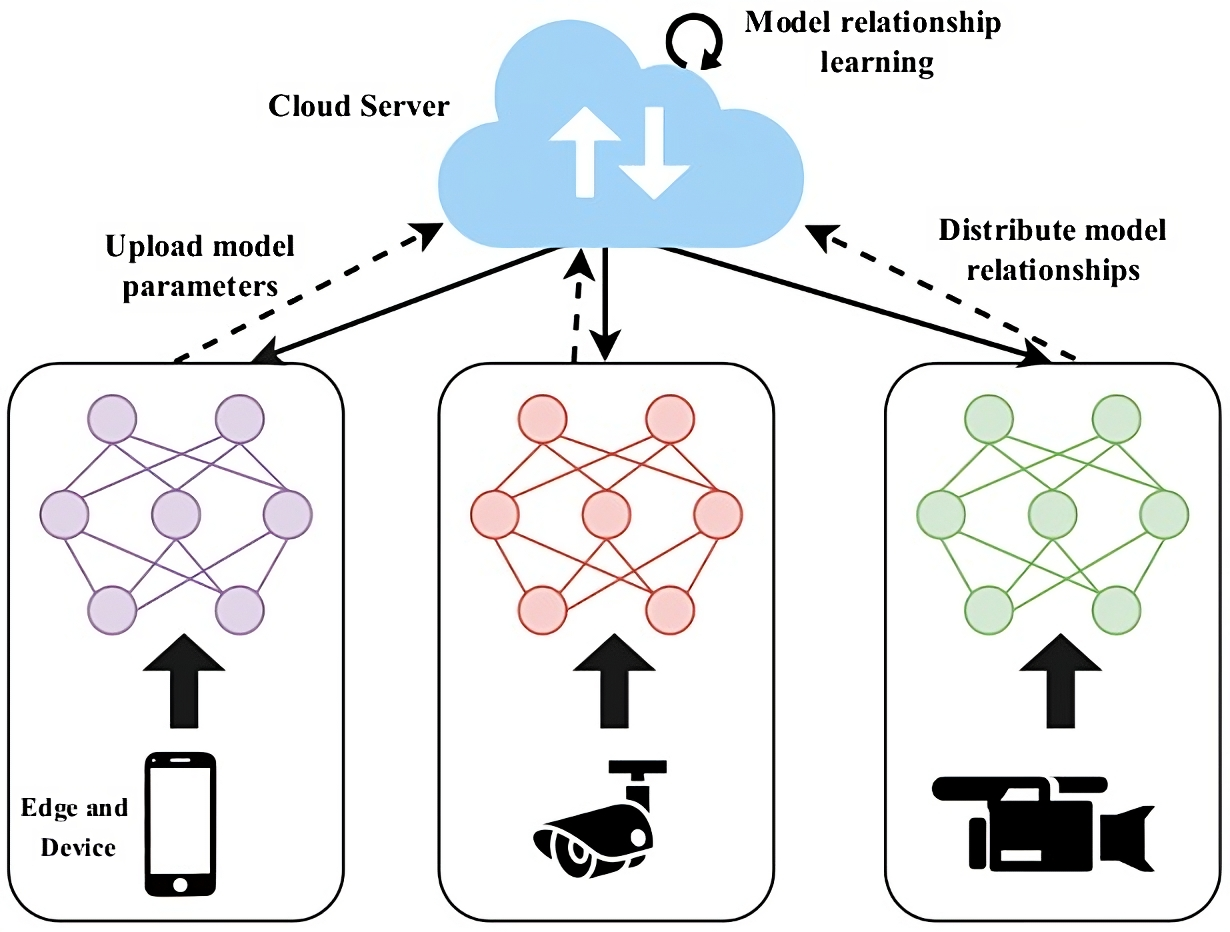}
    \caption{Multitask FL in the healthcare sector~\cite{wu2020personalized}.}
    \label{fig-6}
\end{figure*}

Multitask Federated Learning (MTFL) enables the training of multiple models in parallel using distributed data, where each model is dedicated to a unique task \cite{ricketts2005access}. These models collaborate to learn from each other, improving overall performance as a collective unit \cite{hosny2018artificial}. MTFL has gained prominence in scenarios involving interrelated activities, where data is dispersed across various devices or nodes \cite{chen2021matching}. By leveraging contextual knowledge from different tasks, MTFL enhances model training precision and overall system performance \cite{rischke2022federated}. This technique is particularly advantageous when data decentralization is preferred or necessary \cite{kumar2021federated}. Figure \ref{fig-6} illustrates the application of MTFL in healthcare, showcasing its functionality within this domain.

In MTFL, models are iteratively trained, with local parameter updates consolidated at a central node \cite{shaheen2022applications}. This allows each device's model to benefit from aggregated parameters, improving training efficacy \cite{basu2021benchmarking}. MTFL optimizes joint learning across multiple tasks in a federated setting and ensures privacy and security while achieving high accuracy across tasks \cite{ji2021emerging}. The objective function minimizes the aggregate loss of functions across tasks and devices, with global parameters updated iteratively \cite{yu2022survey}.

\subsubsection{Federated Meta Learning}

Federated Meta-Learning (FML) aims to develop models capable of swiftly adapting to new challenges by leveraging decentralized data sources and prior knowledge from diverse tasks \cite{achituve2021personalized, yu2020salvaging}. In FML, models are iteratively trained on multiple tasks, adjusting their parameters based on acquired knowledge, and then applied to new tasks with revised parameters \cite{fallah2020personalized, wang2022graphfl}.

This approach offers an expedited adaptation to novel tasks without the need for retraining from scratch, saving time and resources \cite{yang2020federated, li2019fair}. Additionally, FML addresses the challenges posed by distributed data, enabling model training across multiple devices without centralized data repositories \cite{dong2022standing, he2020fedml, zhang2020adaptive}. FML optimizes meta-models concurrently across devices, fine-tuning on local tasks while utilizing shared meta-models to acquire task learning knowledge \cite{wang2021review, sarma2021federated}. This optimization facilitates rapid adaptation to new tasks with limited data, ensuring data confidentiality while minimizing meta-model loss across validation tasks \cite{guo2020feel, dinh2022new}.

\begin{figure*}[ht]
    \centering
\includegraphics[width=110mm,height=85mm]{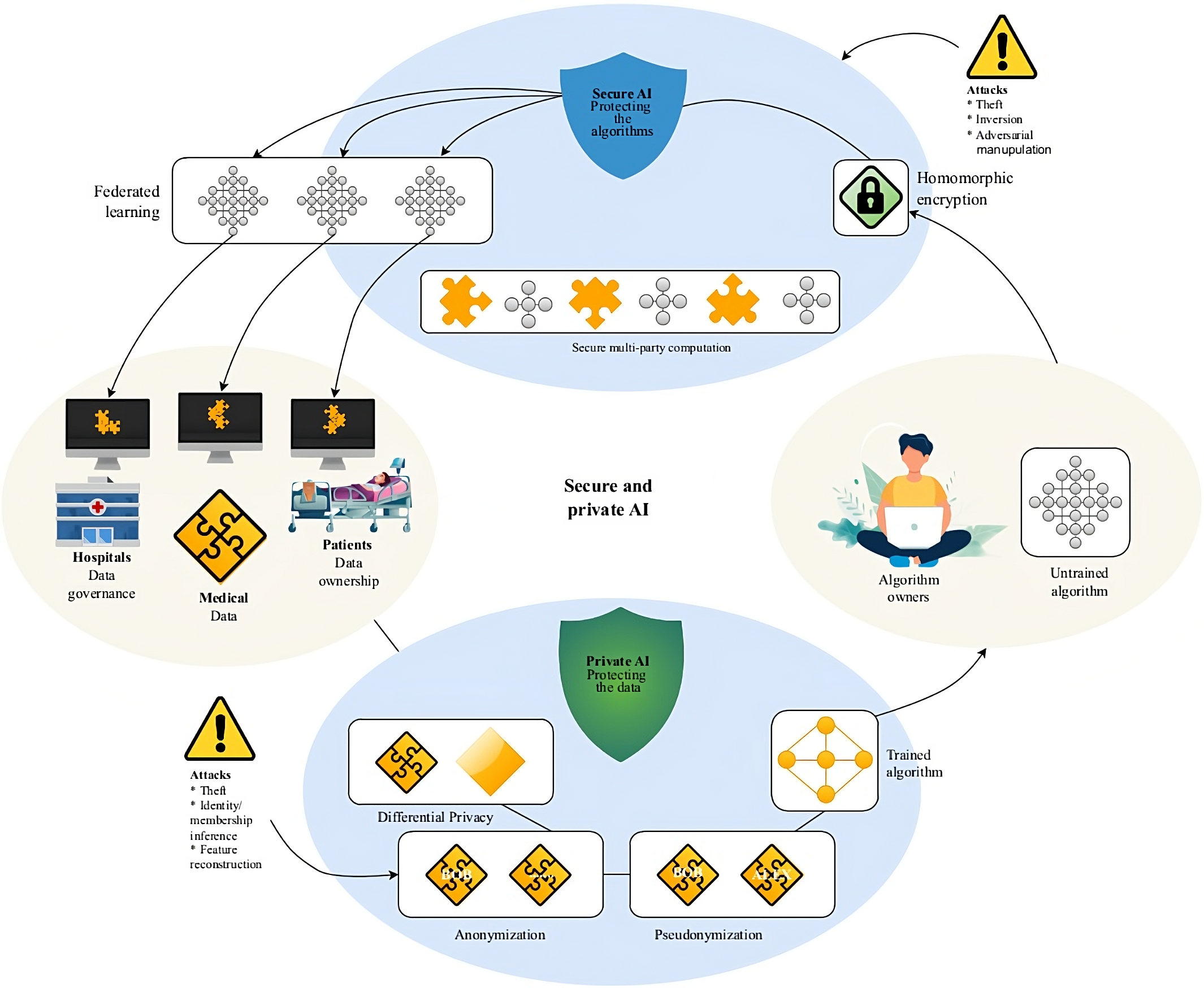}
    \caption{Federated Meta Learning in health care sectors~\cite{chen2018federated}.}
    \label{FML}
\end{figure*}

Moreover, FML trains task-agnostic models with adaptability and versatility, efficiently aggregating decentralized data to develop models capable of performing well across diverse tasks or domains \cite{zhong2022feddar, khan20206g}. Figure \ref{FML} illustrates FML's application in healthcare, demonstrating its functionality within this domain.

\section{Model Misconduct}\label{4}

The present study delves into the phenomenon of model misconduct in the context of FL in healthcare. Specifically, the term "model misconduct" is defined as the deliberate or inadvertent introduction of errors, bias, or other inaccuracies into the ML models developed through the FL approach. This phenomenon has been previously discussed in the literature \cite{papadopoulos2021privacy}. The phenomenon of divergent model performance, whether deliberate or inadvertent, can be attributed to many factors, including dissimilarities in data quality or variations in the selection of algorithms or parameters during the training phase. The occurrence of model misconduct has been identified as a potential source of concern in the medical field, as it may result in erroneous diagnoses, suboptimal treatment outcomes, and other unfavorable consequences for patients. Identifying and remedying model misconduct in FL within the healthcare domain is paramount to guarantee the developed models' precision and impartiality.

\subsection {Data leakage}

FL allows organizations to collaborate on model development without sharing sensitive data, ensuring privacy by enabling local training and aggregating model updates instead of raw data, thus maintaining confidentiality \cite{dhiman2022federated}. The healthcare industry is attracted to FL for its ability to protect highly sensitive patient data. However, it's essential to recognize the risk of data leakage in FL, which could compromise patient privacy and violate data protection regulations \cite{topaloglu2021pursuit}. 

For instance, Chang et al. address data leakage attacks in FL by proposing two defence methods: gradient sparsification and pseudo-gradient techniques. Both methods use cosine similarity to measure gradient differences, and disguise uploaded gradients, protecting privacy without sacrificing model accuracy. Extensive experiments show that these methods effectively safeguard private data while maintaining model performance. However, the effectiveness of these defence methods may vary with different data characteristics and training scenarios, requiring careful parameter tuning~\cite{Chang2024gradient}.

Data leakage, characterized by inadvertent or malicious disclosure of sensitive data to unauthorized parties, poses a significant threat in FL. It can occur during data transmission if communication channels are insecure or if transmission software or hardware vulnerabilities exist. Additionally, insecure storage of data on servers or devices can lead to data leakage if compromised by hackers or malicious actors. Flaws in FL algorithms may also contribute to data leakage if they inadequately protect patient privacy.

To mitigate data leakage risks in healthcare-related FL, organizations should implement security measures such as encryption, authentication protocols, secure storage devices, and privacy-preserving mechanisms in algorithms. Regular monitoring for vulnerabilities and prompt response are crucial to ensuring patient data remains secure throughout the FL process \cite{wang2019adaptive}.

\subsection{Model poisoning}

Model poisoning is a cyber-attack that poses a significant threat to the accuracy and reliability of ML models, including those utilized in FL \cite{ma2022shieldfl}. In healthcare, model poisoning attacks can have severe implications, jeopardizing patient safety and privacy. This type of attack involves an attacker intentionally introducing malicious data into the training dataset to manipulate the behavior of the ML model. The injected data is carefully crafted to bias the model in favor of the attacker, leading to inaccurate or unreliable predictions. By tampering with the training data, the attacker aims to compromise the integrity and effectiveness of the model, potentially causing harm in real-world applications \cite{yang2023review}. Detecting model poisoning attacks in FL can be particularly challenging due to the distributed nature of the training data across multiple healthcare organizations, making it less visible to a central authority \cite{bhagoji2018model}.

\subsection{Model Stealing}

Model stealing attacks involve an attacker replicating an ML model by analyzing its outputs through queries and capturing responses. In FL, this poses a risk for healthcare due to the distributed data across multiple organizations, limiting central authority. Protecting against model stealing requires robust security measures to safeguard sensitive healthcare data and prevent unauthorized model replication \cite{gadekallu2021federated}. 

Sun et al. propose an intrusion detection system (IDS) using Particle Swarm Optimization (PSO) and AdaBoost to detect malware in health app platforms. Using the NSL-KDD dataset, PSO selects 12 relevant features, and AdaBoost achieves a recall value of 0.97. Integrating ML-IDSs into health apps enhances patient care, reduces costs, and secures medical data. This study highlights the potential of ML-IDSs in improving the security of medical IoT devices within the Internet of Medical Things (IoMT)~\cite{sun2024optimized}. However, the approach relies on high-quality input data and assumes the availability of relevant data, which may not always be available. Its dependence on selected features may limit its application where features do not fully capture attack characteristics. 

Differential privacy is a technique that can help mitigate the risk of model stealing by adding random noise to the training data to obscure individual data points. This can make it more difficult for attackers to reconstruct the original data set and reverse-engineer the ML model.

Model watermarking involves adding a unique identifier to the ML model that can be used to detect whether it has been stolen. This can be done by modifying the weights or structure of the model in a way that is imperceptible to its performance but leaves a distinctive pattern that the original owner can recognize. In addition to these techniques, healthcare organizations must implement strong security measures to protect their ML models from unauthorized access and theft. This can include using strong encryption and authentication protocols for data transmission, regularly monitoring the system for potential security vulnerabilities, and promptly addressing any issues.

\subsection {Biased Models}

Biased ML models present a significant risk within FL for healthcare, potentially leading to incorrect diagnoses and inappropriate treatments \cite{pandl2022reward}. 
Kairouz, Peter, et al. scrutinize biased models in FL for healthcare, proposing mitigation strategies and assessing their impact on model performance. Their study delves into various sources of bias and explores methods to address and mitigate bias within the healthcare-focused FL framework \cite{kairouz2021advances}. Inadequate or biased training data, decentralized data distribution, and algorithmic limitations can contribute to bias in ML models. Effectively ensuring representativeness and mitigating bias within FL for healthcare require robust strategies tailored to the decentralized data landscape. Additionally, addressing algorithmic biases is crucial for maintaining model fairness and accuracy in healthcare FL scenarios.

Zhang et al. (2024) utilized robust optimization to enhance FL models, ensuring their consistent performance and fairness across diverse demographics. Their study introduced a unified framework aimed at healthcare institutions, incorporating various fairness metrics and implementing this through an efficient collaborative ML approach. The proposed framework, Federated Learning with Unified Fairness Objective (FedUFO), was tested across four digital medical scenarios in a federated context, demonstrating its ability to preserve model performance while improving fairness. The federated results, particularly for the COVID-19 dataset, showed an overall accuracy of 66.24\%, which underscores the framework's effectiveness in real-world settings~\cite{zhang2024unified}. However, while the framework aims to address and unify diverse levels of fairness in FL, the dependence on collaborative participation from multiple healthcare institutions might introduce variability in adherence to the fairness metrics, potentially impacting the overall efficacy of the model.

\subsection{Incentive Misalignment}

In FL for healthcare, incentive misalignment poses a significant challenge, potentially leading to organizations or individuals prioritizing their interests over the goal of enhancing patient outcomes \cite{zeng2021comprehensive}. Multiple organizations involved in FL for healthcare may have divergent goals and objectives, complicating collaboration and data sharing \cite{dang2020building}. For instance, hospitals may prioritize patient outcomes, while pharmaceutical companies may focus on drug development. Similarly, individuals within these organizations may harbor distinct incentives; for example, physicians may prioritize patient privacy, while researchers may prioritize academic publication. Establishing clear incentives and rewards, such as financial incentives or recognition, can foster collaboration and mitigate the effects of incentive misalignment in FL for healthcare.

\section{Misconduct Threat Models and Adversarial Goals}\label{5}

When multiple devices collaborate to construct a predictive model without disclosing private data, it is referred to as FL \cite{liu2020systematic}. Given the varied trust levels of the involved devices and the potentially sensitive nature of the data, ensuring the security and privacy of FL is paramount \cite{zhao2023privacy}.

\noindent
Both homogeneous and heterogeneous FL present unique opportunities for malicious activity and have diverse adversarial objectives. In a homogeneous FL scenario, where all devices share a similar data distribution, adversaries aim to reduce the model's reliability by interfering with individual device updates. This can be achieved through model poisoning attacks or tampering with the training data \cite{chiu2022application}.

Data manipulation is a prevalent threat model in homogeneous FL, where adversaries can influence the global model by tampering with training data at the local level. For instance, injecting hostile examples into the training data by a rogue device can lead to inaccurate predictions by the global model \cite{ross2023best}. Another security risk involves injecting noisy gradients into the gradients sent by each device to the centralized server during FL training. This malicious action can degrade the precision of the global model \cite{jeong2018communication}. Homogeneous FL also faces threats from model poisoning attacks, where adversaries adjust their local updates to undermine the global model \cite{zhang2015health}. For example, a malicious entity may tamper with the model's parameters to introduce bias, resulting in incorrect predictions in certain cases. In contrast, participating devices in heterogeneous FL have diverse data distributions, and adversaries may have varied motivations \cite{sittig2012survey}. For instance, a device with a skewed data distribution may attempt to incorporate that bias into the overall model. Similarly, a device may choose to conceal certain data by either omitting its input or introducing intentional faults. Ensuring the security of individual data is a significant challenge in FL across different types of devices \cite{revere2007understanding}. Given the potentially sensitive nature of the collected data, some devices may be reluctant to share it. To address this, certain devices may opt to encrypt their local updates, rendering them unreadable by the central server. However, inadvertent or intentional errors introduced into the encrypted updates by the device may adversely impact the overall accuracy of the model \cite{dai2022comprehensive}. Table \ref{table-4} displays the Misconduct Threat Model along with its adversarial goals identified in various research studies.

\begin{table*}[]
\caption{Misconduct threat model and adversarial goals of FL in healthcare.}
\label{table-4}
\scalebox{.92}{
\begin{tabular}{|c|c|c|}
\hline
\textbf{\begin{tabular}[c]{@{}c@{}}Misconduct Threat Model\end{tabular}} & \textbf{Adversarial Goals} & \textbf{Study} \\ \hline
Data poisoning & \begin{tabular}[c]{@{}c@{}}Inject biased/corrupted data or modify existing data to mislead the learning process \\ or infer patient identities.\end{tabular} & \begin{tabular}[c]{@{}c@{}} \cite{ mammen2021federated}\end{tabular} \\ \hline
Model inversion & \begin{tabular}[c]{@{}c@{}}Infer sensitive patient information by exploiting system   vulnerabilities, \\ compromising privacy and confidentiality.\end{tabular} & \begin{tabular}[c]{@{}c@{}} \cite{islam2022privacy}\end{tabular} \\ \hline
Membership inference & \begin{tabular}[c]{@{}c@{}}Determine if a patient's data was used in the learning process, violating  \\ privacy and confidentiality.\end{tabular} & \begin{tabular}[c]{@{}c@{}} \cite{loftus2022federated}\end{tabular} \\ \hline
Byzantine attacks & \begin{tabular}[c]{@{}c@{}}Disrupt the learning process by sending incorrect or malicious updates, leading \\ to incorrect model outputs or system failure.\end{tabular} & \begin{tabular}[c]{@{}c@{}} \cite{ xia2021survey}\end{tabular} \\ \hline
\end{tabular}}
\end{table*}

\subsection{Data Disruption}

FL operates on a decentralized data distribution model, where data is dispersed across multiple devices or participants instead of being centralized in a single location, as seen in traditional ML methods. This distribution offers various advantages. Firstly, it enhances data privacy by allowing sensitive information to remain on individual devices, thus safeguarding it against unauthorized access. Secondly, decentralization ensures system resilience and security by eliminating a single point of failure. Lastly, FL exhibits scalability, as it can efficiently handle large datasets without necessitating centralized data storage solutions.

\begin{table*}[]
\centering
\caption{Summary of public medical datasets used in recent FL studies for medical algorithm benchmarking.}
\label{tabledata}
\scalebox{.88}{
\begin{tabular}{|l|l|l|}
\hline
\multicolumn{1}{|c|}{\textbf{Dataset}} & \multicolumn{1}{c|}{\textbf{Description}} & \multicolumn{1}{c|}{\textbf{Study}} \\ \hline
MIMIC-III & \begin{tabular}[c]{@{}l@{}}A publicly   available dataset containing de-identified electronic health records of over  \\ 40,000 critical care patients from a US hospital system.\end{tabular} & \begin{tabular}[c]{@{}l@{}} \cite{vaid2021federated}\end{tabular} \\ \hline
NIH Chest X-ray & \begin{tabular}[c]{@{}l@{}}A dataset of chest radiographs labeled for the presence of pneumonia, tuberculosis, \\ and other diseases.\end{tabular} & \begin{tabular}[c]{@{}l@{}} \cite{oloko2020tuberculosis} \end{tabular} \\\hline
BraTS & \begin{tabular}[c]{@{}l@{}}A dataset for brain tumor segmentation that includes multimodal magnetic \\ resonance imaging scans of brain tumors.\end{tabular} & \begin{tabular}[c]{@{}l@{}} \cite{li2019privacy} \\  \\ \\ \end{tabular} \\ \hline
Criteo & \begin{tabular}[c]{@{}l@{}}A dataset used in advertising for predicting click-through rates on ads. \\ Used as a proxy dataset for predicting mortality in medical applications.\end{tabular} & \begin{tabular}[c]{@{}l@{}} \cite{vaid2021federated} \end{tabular} \\ \hline
\end{tabular}}
\end{table*}

In FL, participants train local models on their respective data and share model parameters with a central aggregator, consolidating these parameters to construct a global prediction model. However, achieving optimal model performance in FL relies heavily on maintaining balanced data distribution among participants. Imbalanced data distribution can hinder model generalization and compromise its accuracy. Therefore, careful data distribution management is crucial to ensure representative sampling and robust model training.

Moreover, FL algorithms must account for the heterogeneity of data distribution across participants. This necessitates the adaptation of algorithms to accommodate varying data distributions, including differences in missing values, outliers, or other anomalies present in the data. By addressing these considerations, FL can effectively harness the collective knowledge of distributed data sources while mitigating the challenges posed by diverse data distributions (see Table \ref{tabledata} for a summary of public medical datasets in recent FL studies applied for algorithm benchmarking).

\subsection{Data Privacy and Protection}

Data privacy has become a major concern in the era of big data. Numerous security measures and encryption methods have been developed to safeguard sensitive information. These security measures rely on the principle that only individuals with secret keys can access sensitive data. In federated networks, medical data must be stored locally or in the cloud. Hospitals may require on-site or remote computing resources and PACS-type software for data preparation and standardization. However, setting up local centers with data centers and hardware, such as graphics processing units (GPUs), presents challenges such as high computing power requirements, compatibility issues with other clients, and the need for high-performance bandwidth and connections between multiple centers. In medical centers, meeting these requirements may not always be feasible due to inadequate internet connectivity and computational resources~\cite{abdel2021federated}.

Additionally, redundant computing infrastructure and data centers must be designed for the entire network to function correctly and prevent data loss. The network should be resilient enough to continue training even if one computational client fails. Federated models must prioritize network robustness to protect the privacy of models and patient data during client changes or fluctuations in data quantity. This ensures data privacy and model integrity regardless of network dynamics. With the widespread adoption of ML, particularly centralized ML, data collection and transfer to a central location are necessary to train effective models. However, this poses the risk of data leakage for private and sensitive data. Thus, a crucial challenge in sharing intelligence is performing ML on private datasets without compromising data privacy. Multi-party ML with privacy protection mechanisms can assist users in jointly learning from each other's data while ensuring the security of their own data. FL is a prominent example that can help address the privacy concerns associated with multi-party computation.

\begin{table*}[]
\centering
\caption{Related research studies for Medical data properties in FL for medical applications, consisting of data partitions, data distribution (i.e., non-IID) characteristics, possible data privacy attacks, and data privacy protections.}
\label{tableIID}
\begin{tabular}{|c|l|}
\hline
\textbf{Property} & \textbf{Study} \\ \hline
Data partitions & \begin{tabular}[c]{@{}l@{}} \cite{sheller2020federated, li2019privacy} \end{tabular} \\ \hline
Data distribution   (non-IID) characteristics & \begin{tabular}[c]{@{}l@{}} \cite{li2022federated}\\ \end{tabular} \\ \hline

Data privacy   protections & \begin{tabular}[c]{@{}l@{}} \cite{li2023review}\end{tabular} \\ \hline
\end{tabular}
\end{table*}

Related research studies have explored critical aspects such as data partitions, non-IID characteristics, potential data privacy attacks, and corresponding data privacy protections, as summarized in Table \ref{tableIID}.

\subsection{Data Privacy Attacks on ML}

ML data privacy attacks involve unauthorized attempts to access, steal, or alter private information used for training or testing ML models. These attacks can seriously impact individuals and organizations reliant on ML systems, compromising data security, integrity, and availability. Examples of ML data privacy assaults include:

\subsubsection{Data Poisoning}

To bias the final model in their favor, an attacker may inject malicious data into a training dataset, a technique known as data poisoning. For instance, an attacker might introduce false data into the training dataset used to train the filter to enhance the chances of their spam messages being classified as legitimate by a spam filter.

\subsubsection{Model Inversion}

It's a technique where an attacker tries to reconstruct private information, such as an individual's medical records, by exploiting the output of an ML model. By repeatedly testing the model with carefully chosen inputs, the attacker can glean insights into the training data used to construct the model. Membership inference, as an attack method, aims to ascertain whether an individual's data was included in the training set of an ML model. Through analysis of the model's responses to different inputs, the attacker can deduce the presence or absence of specific data, posing risks to privacy and anonymity.

\subsubsection{Model Extraction}

In this method, the attacker seeks to pilfer an ML model by interrogating it with carefully crafted inputs and leveraging the responses to replicate the model. This could pose significant risks, particularly if the model contains sensitive or confidential information.

\subsubsection{Membership inference attacks}

Membership inference attacks seek to ascertain whether a particular individual's data was part of the training dataset for an ML model \cite{bogdanova2020risk}. To carry out these attacks, one needs to train a separate ML model that can identify if a specific data point contributed to the target model's training process \cite{truhn2022encrypted}. Such attacks exploit the use of personal data in the model's training phase to deduce private information about individuals \cite{zhang2021survey}.

\subsubsection{Adversarial examples}

Inputs that have been altered in such a way as to cause an ML model to generate an inaccurate prediction are referred to as adversarial examples \cite{qayyum2022making}. For these assaults to succeed, imperceptible perturbations are first added to the data being used as input. This causes the model to classify the data incorrectly. The reaction of the ML model to the adversarial example can be used to infer information about the original input, which enables sensitive information to be extracted from ML models and utilized for malicious purposes.

\subsubsection{Duties and skills of an attacker}

In ML privacy attacks, perpetrators may adopt various personas and utilize diverse tools to compromise the privacy of individuals or organizations. Attackers can eavesdrop on data being transmitted between parties to gather sensitive information, either by intercepting data in transit or exploiting compromised devices. They may also attempt to obtain the model weights or reverse-engineer the technique to acquire the ML models used by an organization. Even without explicit data access, attackers may try to determine if specific individuals' data was used in training ML models~\cite{dang2021future}. Additionally, based on the outcomes of ML models, attackers may reverse-engineer input data samples, potentially revealing sensitive information. Some attackers create adversarial examples, crafted input data intended to deceive ML models, altering outputs or extracting private data. Using differential privacy methods, attackers may attempt to infer sensitive information from anonymized datasets. These are just a few tactics attackers may use to compromise ML system privacy. Organizations must be vigilant and take measures to protect their data and models. Safeguarding training and evaluation data through encryption and access restrictions is crucial to defend against ML data privacy attacks. Thorough planning and testing of ML models are also essential to identify and mitigate potential vulnerabilities.

\subsection{Data Privacy Protection for Federated Learning}

FL in ML enables collaborative model development among multiple users without sharing sensitive raw data, ensuring privacy and cooperation. Participants train a model on their local devices using their own data, and only model updates are sent to a shared server, preserving data privacy and facilitating collaborative model development. This strategy, keeping data on local devices or servers, helps protect the privacy of sensitive data \cite{rahman2020secure,gadekallu2021federated}. Data privacy protection in FL involves a range of strategies and procedures to prevent the disclosure of sensitive information. One key method is differential privacy, which obscures sensitive data to prevent attackers from deducing it, thereby safeguarding private information \cite{shaik2022fedstack}. Secure aggregation enables parties to combine model updates without revealing individual contributions, ensuring privacy while facilitating collaboration \cite{yu2021federated}. FTL improves model precision by leveraging pre-trained models, reducing the need to distribute sensitive data among parties. Privacy-preserving ML algorithms, such as secure decision trees and k-means clustering, protect data confidentiality effectively. Data access control manages who can access what data, restricting access to private information and ensuring only authorized users can access it \cite{le2021fedxgboost}.

\section{Federated Learning for Healthcare Applications}\label{6}

\subsection{COVID-19 Detection}

FL has been effectively employed in the context of COVID-19 to aid in virus detection using wearable technology and various medical tools \cite{naz2022comprehensive,chowdhury2022federated}. For instance, smartwatches and activity trackers can monitor COVID-19 symptoms by gathering data on temperature, heart rate, and other vital indicators, facilitating early detection of the virus. However, collecting such vast amounts of information from millions of devices globally poses challenges and raises concerns regarding centralized databases, particularly regarding privacy and security. FL addresses these challenges by allowing data to be stored locally on devices and models to be trained in a decentralized manner. This approach shows promise in leveraging COVID-19 EHR data to develop accurate predictive models while safeguarding patient privacy \cite{vaid2021federated}. Another application is the detection of COVID-19 through X-ray imaging, a commonly used method in diagnosing respiratory disorders. Salam et al. developed ML models, including an FL model, using chest X-ray (CXR) images from COVID-19 patients, showcasing the potential of FL in this domain \cite{abdul2021covid}. By keeping the images locally stored at hospitals and clinics, FL maintains privacy and security while enabling model training on a sizable dataset \cite{feki2021federated}. Furthermore, FL can be utilized in COVID-19 detection through CT scans, where models can be trained on a large dataset of CT images to identify distinctive patterns associated with the virus. The approach proposed by Rajesh et al., combining blockchain technology and FL for joint model training, ensures anonymity while enhancing COVID-19 detection accuracy \cite{kumar2021blockchain}. Overall, FL offers a valuable method for enhancing COVID-19 detection accuracy and speed while preserving data confidentiality and security amidst the increasing utilization of wearable technology and medical equipment.

\subsection{Breast Density Classification:}

Classifying breast density is crucial for assessing a patient's risk of developing breast cancer. Modern technology called FL offers a promising avenue to enhance breast density classification. With FL, several parties can collaborate decentralized while maintaining the confidentiality and privacy of their data \cite{crimi2022brainlesion}. To train an ML model for classifying breast density, several hospitals or imaging facilities may pool their data while keeping the identities of their separate datasets a secret. The lack of extensive, high-quality datasets that adequately represent the broad community of mammography-using women is one of the main obstacles to breast density classification. Using an FL method, the open-source software MammoDL employs the U-Net DL architecture to quantitatively analyze the density and complexity of breast tissue from mammograms \cite{katti2022mammodl}. By combining smaller datasets from several institutions into a more extensive, more diversified dataset, FL can aid in overcoming this difficulty \cite{roth2020federated}. This additional diversity can enhance the ML model's performance and lower the danger of overfitting, which happens when a model is trained too closely on a particular dataset and struggles to generalize to new data. Additionally, FL may help with privacy issues. All data must be centralized in a standard ML scenario, making it susceptible to hackers and data breaches. In FL, the raw data is still stored on the servers of the individual universities and is only utilized to update models. Patient privacy is safeguarded, and the risk of data leakage is reduced.

\subsection{Healthcare Monitoring}

FL promises a transformative impact on healthcare monitoring by enabling collaboration among academia and healthcare professionals \cite{nguyen2022federated}. By pooling data from diverse sources like medical devices, patient records, and wearable technology, FL facilitates the development of accurate ML models to forecast patient health outcomes \cite{rieke2020future}. These models predict disease susceptibility, readmission risks, and medication responses using data from electronic health records, medical devices, and patient-generated data. Wearable devices, such as fitness trackers, can remotely monitor vital signs, activity levels, and sleep patterns, alerting healthcare providers to any deviations. Proposed frameworks, like edge-assisted data analytics, employ FL to locally retrain ML models using user-generated data while preserving privacy \cite{hakak2020framework}. FL ensures patient data privacy by retaining data locally and encrypting it before transmission to a central server, reducing the risk of data breaches and unauthorized access \cite{shaik2022fedstack}. Overall, FL holds the potential to revolutionize healthcare monitoring by improving forecast accuracy, providing personalized care, and safeguarding patient privacy.

Dheeba et al. (2024) explore a decision support system for heart disease prognosis, deploying it on both cloud-native and edge-optimized models. The system, named ClassifyIT, employs a custom neural network architecture called IPANN, a feature selector named MIST-CC, and a regularizer named STIR. ClassifyIT achieved an accuracy of 87.16\% on the Cleveland dataset, outperforming a regular deep network’s 78.80\%. Adding MIST-CC improved the accuracy to 81.97\%, and including STIR further enhanced it to 85.54\%~\cite{dheeba2024heart}.
The study introduces two deployment strategies: cloud-native and edge-optimized. The cloud-native model offers a centralized, scalable solution, while the edge-optimized model improves scalability by shifting computation to user devices. The ML pipeline is further refined using FL, which promotes localization and collaborative learning. The edge-optimized architecture ensures better scalability and adaptability for real-world applications. However, potential limitations of the proposed methods include data availability and integration across cloud and edge platforms. Moreover, the system's reliance on FL demands robust infrastructure and coordination among multiple devices. Additionally, it is unclear whether their proposed approaches are user-friendly.

\subsection{ Medical Imaging}

FL in healthcare, particularly in medical imaging, involves using ML techniques to process medical images decentralized \cite{kaissis2020secure, adnan2022federated}. This approach is vital for healthcare because medical imagery is crucial in diagnosing and treating patients. Medical imaging data is typically collected from various sources like imaging facilities, hospitals, and clinics, each with unique datasets. FL allows these entities to collaborate in training global models, which improves disease identification and treatment effectiveness over time as more data becomes available \cite{rieke2020future}. FL addresses privacy concerns inherent in centralized data systems by keeping data secure and private while leveraging collective knowledge from multiple organizations to enhance model development \cite{darzidehkalani2022federated}. FL algorithms, such as Federated Averaging (FedAvg), enable real-time collaboration among organizations, accelerating model development and disease diagnosis \cite{mcmahan2017communication}. FL also addresses data imbalance issues common in medical imaging by pooling data from diverse sources, resulting in more robust models and accurate diagnoses \cite{pan2019improving, chang2018distributed}. Overall, FL promises to improve patient outcomes and reduce healthcare costs in medical imaging applications. Various publications have explored the applications of FL in medical contexts, summarizing their findings, methodologies, and outcomes, as shown in Table \ref{Table:app}.

\begin{table*}[]
\caption{Summary of FL publications applied in medical applications.}
\label{Table:app}
\scalebox{1}{
\begin{tabular}{|l|l|l|l|l|}
\hline
\textbf{ML Task} & \textbf{Clinical Tasks} & \textbf{Medical Input Data} & \textbf{Model Architecture} & \textbf{Related FL Study} \\ \hline
Classification & \begin{tabular}[c]{@{}l@{}}COVID-19 \\ Diagnosis\end{tabular} & \begin{tabular}[c]{@{}l@{}}Chest X-ray \\ images\end{tabular} & CNN, ResNet-18 & \begin{tabular}[c]{@{}l@{}} \cite{de2022fully}\end{tabular} \\ \hline
Classification & \begin{tabular}[c]{@{}l@{}}Diabetic Retinopathy \\ Diagnosis\end{tabular} & Fundus images & CNN, MobileNet-v2 & \begin{tabular}[c]{@{}l@{}} \cite{teo2022developments}\end{tabular} \\ \hline
Segmentation & \begin{tabular}[c]{@{}l@{}}Brain Tumor \\ Segmentation\end{tabular} & MRI scans & CNN, U-Net & \begin{tabular}[c]{@{}l@{}} \cite{foley2022openfl}\end{tabular} \\ \hline
Segmentation & \begin{tabular}[c]{@{}l@{}}Cardiac MRI \\ Segmentation\end{tabular} & MRI scans & CNN, U-Net & \begin{tabular}[c]{@{}l@{}} \cite{kwan2021artificial}\end{tabular} \\ \hline
Segmentation & \begin{tabular}[c]{@{}l@{}}Lung Tumor \\ Segmentation\end{tabular} & CT scans & CNN, U-Net & \begin{tabular}[c]{@{}l@{}} \cite{nazir2023federated}\end{tabular} \\ \hline
Regression & \begin{tabular}[c]{@{}l@{}}Mortality \\ Prediction\end{tabular} & \begin{tabular}[c]{@{}l@{}}Electronic Health \\ Records\end{tabular} & LSTM, MLP & \begin{tabular}[c]{@{}l@{}} \cite{rajendran2023data}\end{tabular} \\ \hline
Regression & \begin{tabular}[c]{@{}l@{}}Blood Pressure \\ Prediction\end{tabular} & PPG signals & LSTM & \begin{tabular}[c]{@{}l@{}} \cite{brophy2021estimation}\end{tabular} \\ \hline
Regression & \begin{tabular}[c]{@{}l@{}}Heart Failure \\ Prediction\end{tabular} & \begin{tabular}[c]{@{}l@{}}Electronic Health \\ Records\end{tabular} & Multi-task LSTM & \begin{tabular}[c]{@{}l@{}} \cite{abdulrahman2020survey}\end{tabular} \\ \hline
\end{tabular}}
\end{table*}

For instance, Joynab et al. (2024) address cervical cancer detection using CNN-based FL architectures, balancing accurate image classification and data privacy across three experimental settings. The proposed system aggregates updates from locally trained models into a global model, achieving test accuracies of 94.36\% in an IID setting and 78.4\% in a non-IID setting~\cite{joynab2024federated}. However, the model's performance varies significantly between IID and non-IID settings, highlighting potential challenges with heterogeneous data. The reliance on FL also requires robust infrastructure and cooperation among institutions, which may not always be feasible. 

Truhn et al. (2024) address privacy concerns in AI training for cancer image analysis using Somewhat-Homomorphically-Encrypted Federated Learning (SHEFL). This method transfers only encrypted weights, avoiding data breaches during model updates. SHEFL was successfully implemented in various cancer image analysis tasks using multicentric datasets. For instance, SHEFL models achieved an 80.32\% Dice score, close to the 81.71\% of FL models~\cite{truhn2024encrypted}. However, the computational overhead of homomorphic encryption can slow down training. Additionally, while SHEFL improves privacy, it does not entirely eliminate risks. The method also depends on secure infrastructure and collaboration between institutions. 

Ahsan et al. (2024) proposed models such as M-VGG16, M-ResNet50, M-ResNet101, and ViT for classifying Monkeypox using image analysis. The M-VGG16 achieved 88\% accuracy on a small dataset and 76-77\% on an imbalanced dataset. The M-ResNet50 performed best with 89\% accuracy on a multiclass dataset. Models trained with Adam surpassed those trained with SGD. The study highlights the use of FL to improve AI-based diagnostic models by enabling collaborative model training without data sharing, thereby preserving patient privacy. The integration of FL demonstrates the potential for developing robust and secure healthcare solutions~\cite{ahsan2024enhancing}. However, the study faces several limitations, including a limited number of Monkeypox images, the need for expert verification, testing on highly imbalanced data, and the development of mobile diagnostic tools.

\subsection{Electronic Health Record}

FL has the potential to enhance clinical decision-making within EHR systems significantly \cite{dang2022federated}. Initially proposed by Google for board question suggestion, FL involves training a global model using data from various sources, including wearable technology, hospital systems, and individual health records \cite{brisimi2018federated}. Each participant's device trains the global model locally, with parameters then aggregated to update the global model. FL addresses the issue of data silos in EHR systems, allowing insights and predictions based on a broader and more diverse dataset without moving the data \cite{alzubi2022cloud}. 

Kuliha et al. (2024) address the significant role of EHRs in healthcare, focusing on the challenges of data management and security through FL and blockchain technology. They tackle issues like motivating participation in FL, ensuring accurate model aggregation, and managing the large volume of EHR data. Their proposed solution integrates blockchain with cloud solutions to enhance interoperability, albeit at the cost of compromising the immutability of EHRs~\cite{kuliha2024secure}. However, the integration of blockchain could lead to potential delays in updating EHRs due to additional validation steps. 

Pan et al. (2024) propose an adaptive FL framework to address data distribution drift across institutions in clinical risk prediction using EHR. This framework separates input features into stable, domain-specific, and conditional-irrelevant parts based on their relationship to clinical outcomes. Evaluated on a large-scale intensive care unit (ICU) dataset, the framework predicts the onset risk of sepsis and acute kidney injury (AKI) more effectively than existing FL baselines, while also being clinically interpretable.

The experimental results demonstrate that the proposed method outperforms other FL approaches, with notable AUROC scores for AKI and sepsis prediction across multiple hospitals. For example, the AUROC for AKI prediction using the proposed method ranged from 0.679 to 0.818 across hospitals, surpassing other methods like APPLE and RMTL. For sepsis prediction, the proposed method achieved AUROC scores from 0.659 to 0.861, again outperforming other techniques~\cite{pan2024adaptive}. However, the study did not mention whether their proposed method can handle the computational complexity or how their proposed method handles the data variability for clinical practices across various institutions

Some propose combining Deep Learning (DL) and blockchain technologies to enhance privacy protection in EHR, using techniques like anomaly detection and cryptography-based FL modules \cite{alzubi2022cloud}. Despite its potential benefits, integrating FL into EHR faces challenges such as data privacy, security, model fairness, accuracy, and legal and ethical concerns. Addressing these obstacles is crucial to ensure patient data confidentiality and privacy while leveraging FL to improve clinical decision-making.

\section{Challenges of Federated Learning}\label{7}

While FL offers the potential for multiple devices to collaborate on training ML models without sharing raw data, it also presents several challenges that need to be addressed \cite{kairouz2021advances}. Figure~\ref{fig88} illustrates a framework of challenges associated with FL. These significant challenges include:

\subsection{Heterogeneity of Data and Devices:
}
The variety of data and devices makes deploying FL effectively in the healthcare industry challenging.

\begin{itemize}
    \item \textbf{Data Heterogeneity:}
    Healthcare data exists in diverse formats such as text, images, and time series, each with unique characteristics requiring specific processing methods for model training. Variations in data quality among devices, influenced by device specifications, sensor accuracy, and patient demographics, pose challenges. Therefore, data pre-processing and standardization are essential to ensure the success of FL.
    
    \item \textbf{Device Heterogeneity:} Healthcare devices vary in hardware and software requirements, network connectivity, and power consumption. Some devices may be resource-constrained and lack computational or memory capabilities to participate in FL. Additionally, disparities in operating systems or programming languages among devices require additional data interoperability and communication efforts. Various strategies have been devised to address this heterogeneity in FL so that healthcare can overcome differences in data and devices.

\item \textbf{Federated Transfer Learning:} This method uses pre-trained models on related data domains to speed up model training on heterogeneous devices. Algorithms for adaptive learning can modify the model parameters by the device's processing power. Employing efficient communication protocols to decrease data transfer and lower device computing load is possible.

\end{itemize}

\subsection{Data privacy and security:}

Even though this strategy has several benefits, including enhanced privacy and lower communication costs, it also presents severe problems for data protection~\cite{truong2021privacy}. The following are a few of the significant issues with data privacy that FL presents:
\begin{itemize}
    \item \textbf{Leakage of data:} In FL, devices distribute updates to the model, but these changes could still include private information about the local data~\cite{jin2021cafe,wei2020framework}. Potentially, attackers can intercept these updates and deduce private information about the local data. Encryption and other privacy-protecting methods must be used to stop data leakage \cite{zhao2020idlg}.
    \item \textbf{Model inversion attacks:} This privacy problem is also possible in FL. These attacks use the model updates to recreate the initial training data utilized by the local devices \cite{phong2017privacy}. It is essential to put privacy-preserving procedures that safeguard the local data to stop these assaults.

    \item \textbf{Attacks using membership inference:} Attackers may be able to identify which device was used to train the model in FL if they have access to certain devices. The device’s user or the local data may be revealed via this information in sensitive ways \cite{nasr2019comprehensive}. Employing acy-preserving methods like differential privacy can stop membership inference attacks \cite{lyu2020threats}.

    \item \textbf{Attacks on the central server:} In FL, the central server gathers model updates from the local devices. An attacker may gain access to all updates and deduce vital information about the local data if the primary server is hacked \cite{li2020learning}. Therefore, it is essential to put in place robust security measures to safeguard the central server.

\end{itemize}

\begin{figure*}[ht]
    \centering
\includegraphics[width=\textwidth]{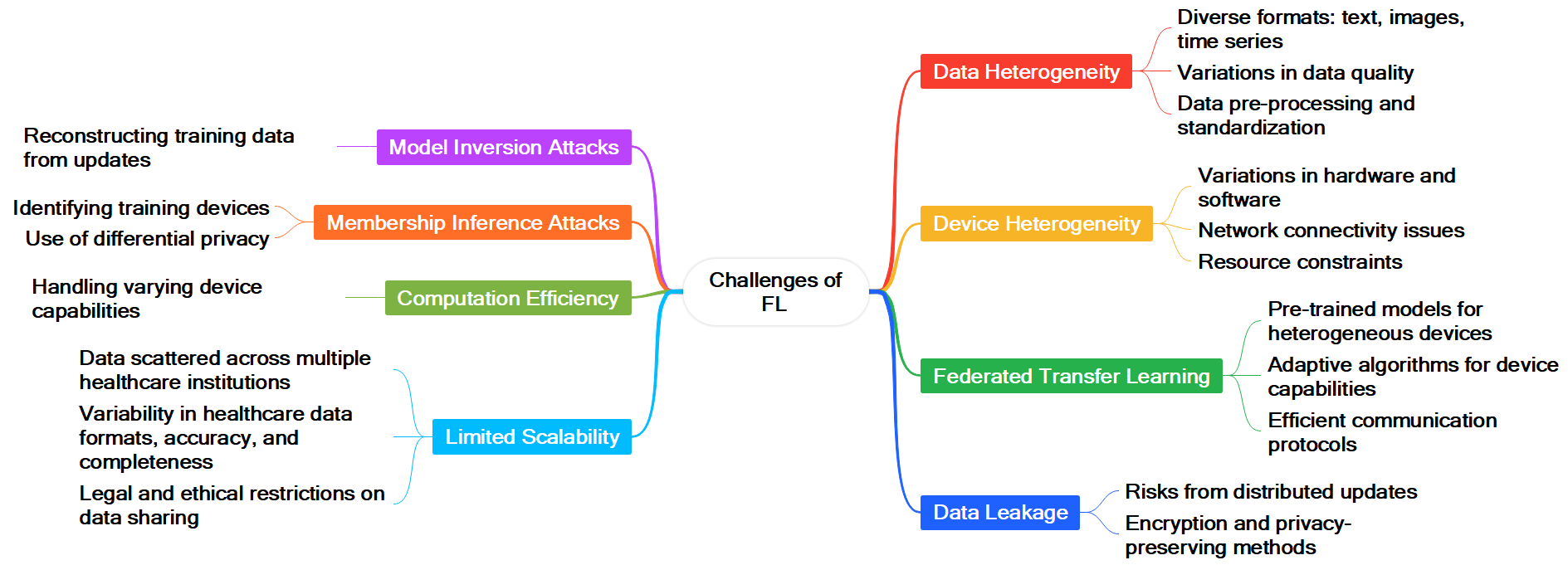}
    \caption{Challenges of FL in healthcare.}
    \label{fig88}
\end{figure*}

Overall, protecting data privacy is a significant difficulty in FL. Overcoming these challenges requires a mix of privacy-preserving strategies, strong security precautions, and rigorous FL system design and implementation. Moreover, Figure \ref{fig88} shows the challenges of FL in different health sectors.

\subsection{ Communication and computation efficiency:
}
Communication effectiveness poses a crucial issue for FL since the data from the participating devices must be sent to the central server for model training securely and promptly~\cite{konevcny2016federated}. The data must be delivered with the least amount of delay possible to guarantee that the models are updated as soon as possible and to safeguard the privacy of the user's data. However, FL also needs help with computing efficiency \cite{diao2020heterofl}. As a result, the model’s performance may vary depending on how powerful the participating devices are. Additionally, the models must be trained using the device's meager resources, which could result in excessive battery usage. FL methods are constantly updated to address these issues \cite{konevcny2016federated}. Adopting model compression techniques to shrink the model’s size may facilitate communication more quickly, which is one way to get around the problem of communication and computation efficiency ~\cite{choi2020communication}. Additionally, data partitioning, selective participation, and model parallelism can be used to compute efficiency.
In conclusion, FL faces substantial communication and computation efficiency issues that must be overcome to increase the technique's overall effectiveness. Researchers and practitioners constantly look for ways to improve communication and computation efficiency to keep FL a viable choice for ML.

\subsection{Handling Non-IID (Independent and Identically Distributed) data:}

Non-IID (Independent and Identically Distributed) data handling is one of the greatest issues with FL~\cite{zhu2021federated}. Traditionally, it has been assumed that the training data in ML is independent and uniformly distributed (IID). However, with FL, the information is gathered from numerous sources, some of which need to be more independent or evenly distributed. The data's non-IID character presents several difficulties in FL \cite{mcmahan2017communication,singh2023fair}. For instance, data distribution may vary significantly between different customers, making it difficult to train a general model that can perform well across all clients. Non-IID data could introduce bias into training, leading to poor performance and negative results~\cite{li2022federated}.

Several methods, including client weighing, data augmentation, and transfer learning, have been suggested to overcome these difficulties. Client weighting involves assigning different weights to individual clients based on the distribution of their data. This approach allows the training process to focus more on clients with data that is more representative, improving the model's overall performance. By introducing noise or making minor adjustments to the current data, data augmentation includes creating fresh data samples that can be used to help make the data more representative. FedProx \cite{li2020federated}, SCAFFOLD \cite{karimireddy2020scaffold}, and FedNova \cite{wang2020tackling} are some studies that have attempted to create efficient FL algorithms for non-IID data. Using a pre-trained model as a starting point for training on the non-IID data is one transfer learning method that can help overcome the challenges posed by non-IID data (see Table \ref{tableNon-IID} for a comparison of data partitioning and non-IID characteristics in FL approaches for healthcare applications).

\begin{table*}[]
\centering
\caption{Comparison of Data Partitioning and Non-IID Characteristics in FL Approaches for Healthcare Applications.}
\label{tableNon-IID}
\begin{tabular}{|c|c|c|c|c|c|}
\hline
\textbf{Study} & \textbf{Data Partition} & \textbf{\begin{tabular}[c]{@{}c@{}}Number \\ of   Nodes\end{tabular}} & \textbf{\begin{tabular}[c]{@{}c@{}}Non-IID \\ Characteristics\end{tabular}} & \textbf{Non-IID  Mitigation} & Study \\ \hline
{[}1{]} & Random & 5-10 & \begin{tabular}[c]{@{}c@{}}Data from different   \\ hospitals\end{tabular} & Distillation-based & \cite{abdelwahab2021concealment} \\ \hline
{[}2{]} & Random & 5-10 & \begin{tabular}[c]{@{}c@{}}Different hospitals and \\ patient demographics\end{tabular} & None &\cite{wang2020mfpc}\\ \hline
{[}3{]} & Random & 5-10 & \begin{tabular}[c]{@{}c@{}}Different hospitals   and \\ imaging modalities\end{tabular} & None & \cite{singh2020optimizing} \\ \hline

{[}4{]} & Stratified & 10 & \begin{tabular}[c]{@{}c@{}}Different hospitals   and \\ patient demographics\end{tabular} & Multi-task learning & \cite{singh2020optimizing} \\ \hline
{[}5{]} & Stratified & 4 & \begin{tabular}[c]{@{}c@{}}Different hospitals   and \\ patient demographics\end{tabular} & Domain adaptation & \cite{hauptmann2020deep} \\ \hline

{[}6{]} & Stratified & 10 & \begin{tabular}[c]{@{}c@{}}Different hospitals   and \\ patient demographics\end{tabular} & \begin{tabular}[c]{@{}c@{}}Federated transfer   \\ learning\end{tabular} & \cite{saffold2021dielectric} \\ \hline
{[}7{]} & Random & 10 & \begin{tabular}[c]{@{}c@{}}Different hospitals   and \\ patient demographics\end{tabular} & \begin{tabular}[c]{@{}c@{}}Local training and   \\ global aggregation\end{tabular} & \cite{ozleyen2021crowd} \\ \hline

{[}8{]} & Random & 5 & \begin{tabular}[c]{@{}c@{}}Different hospitals   and \\ patient demographics\end{tabular} & \begin{tabular}[c]{@{}c@{}}Multi-level federated \\ learning\end{tabular} & \cite{song2021power} \\ \hline
{[}9{]} & Stratified & 4 & \begin{tabular}[c]{@{}c@{}}Different hospitals   and \\ patient demographics\end{tabular} & Differential privacy & \cite{boob2020feasible} \\ \hline
{[}10{]} & Stratified & 4-10 & \begin{tabular}[c]{@{}c@{}}Different hospitals   and \\ patient demographics\end{tabular} & Meta-learning & \cite{nayak2020msit} \\ \hline
\end{tabular}
\end{table*}
 
In conclusion, one of the main difficulties in FL is processing non-IID data. This problem has been addressed using several strategies, such as client weighting, data augmentation, and transfer learning. More study is necessary to create more practical ways for managing non-IID data in FL.

\subsection{Data Bias}

Data bias is the term used to describe a situation in which the data used to train the model is inaccurate or slanted in some way \cite{wang2020tackling}. This may occur when specific servers or devices have more data than others, when some servers have different types of data than others, or when the data is not randomly sampled. As a result, the model might develop biases in favour of the training data, negatively affecting how well it performed on new, untrained data. Data bias can occur in FL in a variety of ways. For instance, if some devices have a different distribution of data or more data than others, the model may become biased towards those devices and perform poorly on the devices with varying allocations of fewer data. Additionally, if the data is skewed or not sampled randomly, the final model might not accurately reflect the entire population. Several strategies can be utilized to solve the problem of data bias in FL. One process is to guarantee that the data's sample size and population representation are random. Using data augmentation techniques to produce additional training data representative of the entire population is an alternative strategy. Additionally, FL algorithms can be created to account for data bias by providing greater weight to devices with fewer data or a different distribution or by utilizing adaptive learning algorithms that modify the model according to the data distribution on each device.

\subsection{Limited scalability}

Limited scalability is one problem with FL in healthcare. Because healthcare data is usually scattered among several hospitals, clinics, and other healthcare providers, organizing the data sharing and aggregation necessary for FL can be difficult. An additional challenge is the unpredictability of medical data. Due to the variety of healthcare data in format, accuracy, and completeness, FL-trained ML models may perform differently. Additionally, the data that can be shared and utilized to train ML models may be restricted due to legal and moral restrictions on healthcare data.
Researchers are looking into several methods to increase the FL in healthcare's capacity for scaling to overcome these issues. One of these is establishing standards and protocols for data sharing and collaboration among healthcare providers. Others include creating new algorithms to handle heterogeneous and distributed data, developing privacy-preserving techniques to safeguard sensitive information, and creating privacy-preserving processes \cite{khan2021federated}. To guarantee that the advantages and hazards of FL are pretty balanced and to build trust in its application in the healthcare industry, stakeholders must also be included, including patients, providers, and policymakers.

\subsection{Lack of interpretability}

 Understanding how a model generates its predictions or judgments is called interpretability. Transparency, accountability, and trust are just a few reasons why it's crucial. Nevertheless, FL's dispersed raw data and centralized server that aggregates model updates make it challenging to understand the whole model. The issue of interpretability in FL can be solved by employing techniques like differential privacy, which adds noise to the data to protect individual privacy while still allowing the model to be properly trained. Using techniques like model distillation is a different tactic. The global model is trained on top of a simpler model utilizing the same data set. Overall, FL presents interpretability challenges, but a number of approaches can be used to address these challenges and enable the development of trustworthy and open models.

 \section{Possible Future Research Directions}\label{8}
 Further research in these areas can help unlock the full potential of FL for improving healthcare outcomes while protecting patient privacy. There are several potential research directions for FL in healthcare, including:
 
\begin{itemize}
    \item \textbf{Developing privacy-preserving techniques:}
    Developing privacy-preserving techniques will be a critical focus of future research in FL in healthcare, as patient data is susceptible and must be protected.
    
    \item \textbf{Model generalization:} ML models developed using FL may be prone to overfitting, which can limit their effectiveness in real-world settings. Improving model generalization in FL can help to ensure that these models are accurate and effective when deployed in healthcare settings. So, to enhance the effectiveness of FL models in real-world healthcare settings, researchers will need to focus on improving model generalization and reducing overfitting.
    
    \item \textbf{Identifying and addressing data biases:} Biases in healthcare data can perpetuate and amplify existing disparities and inequities. Identifying and addressing data biases in FL ensures that these models are fair and equitable and can help reduce healthcare disparities. So, addressing data biases can be a focused area of future research in FL in health care in the future.
    
    \item \textbf{Developing new architectures for healthcare-specific tasks:} Healthcare has unique data and task requirements that may require specialized architectures for FL models. So new architectures and models need to be developed for better performance.
    
    \item \textbf{Combining FL with other technologies:} FL can be combined with other technologies, such as blockchain and edge computing, to create new healthcare solutions. Combining FL with future technologies can help address some unique challenges and opportunities in healthcare.
    
    \item \textbf{Applying FL in low-resource settings:} FL can be a potent tool in low-resource settings, where there may be limited access to data and computing resources. It also can be an important area of future research in FL in healthcare, as it can potentially improve healthcare outcomes in underserved areas.
 
\end{itemize}

Overall, there are many exciting opportunities for future research in FL in healthcare. As the technology continues to evolve, it will be necessary for researchers to address these challenges and develop new techniques and tools to enable the widespread adoption of FL in healthcare.

\section{Conclusion}\label{9}
In this study, we explore the application of Federated Learning (FL) in healthcare, highlighting its potential to enhance privacy and security in handling sensitive medical data. Our findings show that FL can significantly mitigate risks such as data leakage and model inversion attacks through techniques like differential privacy and secure aggregation. However, challenges remain, particularly in managing data and device heterogeneity, ensuring communication and computation efficiency, and addressing the non-IID nature of healthcare data. These challenges complicate the deployment of FL and pose significant technical, ethical, and logistical barriers to its broader adoption. Therefore, Future work should focus on developing robust, scalable algorithms that can efficiently handle heterogeneous and distributed datasets while maintaining patient privacy and data security. Research should also focus on the interpretability and fairness of FL models, ensuring they are accessible and equitable across diverse patient demographics. Addressing these issues will improve the operational efficiency of FL and build trust among stakeholders, encouraging its integration into mainstream healthcare practices. The potential impact of successfully integrating FL in healthcare is profound, promising to revolutionize patient care by enabling more personalized and preventive medicine while securing patient privacy.

\bibliographystyle{unsrt}  
\bibliography{main}

\end{document}